\documentclass{article}

\usepackage{arxiv}

\usepackage{xspace}
\usepackage{maltese}
\usepackage{amssymb}
\usepackage{amsmath}
\usepackage[normalem]{ulem}
\usepackage{subcaption}
\usepackage{hyperref}
\usepackage[usenames,dvipsnames,svgnames,table]{xcolor}
\usepackage{paralist}
\usepackage{multicol}
\usepackage{graphicx}

\newcommand{\vjagg}{\textit{vja\malteseg\malteseg}\@\xspace}
\newcommand{\Vjagg}{\textit{Vja\malteseg\malteseg}\@\xspace}
\newcommand{\VJAGG}{\textit{VJA\malteseG\malteseG}\@\xspace}
\newcommand{\hav}{hav}
\newcommand{\speed}[1]{v\left( #1 \right)}
\newcommand{\ie}{i.e.\@\xspace}

\newcommand{\foottext}[1]{{\let\thefootnote\relax\footnote{#1}}}

\title{\VJAGG\ -- A Thick-Client Smart-Phone Journey Detection Algorithm}

\author{
  Michael P. J. Camilleri \\
  School of Informatics\\
  University of Edinburgh, UK \\
   \And
  Adrian Muscat \\
  Faculty of ICT \\
  University of Malta, Malta \\
   \And
  Victor Buttigieg \\
  Faculty of ICT \\
  University of Malta, Malta \\
   \And
  Maria Attard \\
  Institute for Climate Change \& Sustainable Development \\
  University of Malta, Malta
}

\begin{document}
\maketitle

\begin{abstract}
In this paper we describe \Vjagg, a battery-aware journey detection algorithm that executes on the mobile device. The algorithm can be embedded in the client app of the transport service provider or in  a general purpose mobility data collector. 
The thick client setup allows the customer/participant to select which journeys are transferred to the server, keeping customers in control of their personal data and encouraging user uptake.
The algorithm is tested in the field and optimised for both accuracy in registering complete journeys and battery power consumption.
Typically the algorithm can run for a full day without the need of recharging and more than 88\% of journeys are correctly detected from origin to destination, whilst 12\% would be missing part of the journey.
\end{abstract}


\foottext{We acknowledge financial support from the Vodafone Malta Foundation, under the ``Connecting for Good'' project. The work was done while Michael P. J. Camilleri was employed with the Faculty of ICT at the University of Malta.}

\section{Introduction}
Typically, transport planners, researchers and service providers rely on infrequent and expensive travel diary surveys to collect information relative to journeys made during the day, their purpose, mode used and other useful information to derive population travel behaviour. Traditional pen and paper travel diary surveys suffer from a number of limitations. From a logistical perspective, surveys entail a determined commitment on behalf of participants to either recall or continuously track their activities and nevertheless, data is typically inaccurate due to errors in human judgement \cite{stopher_2015, geurs_2015}.  Fortunately, the availability of low-cost GPS trackers and the subsequent boom in the smartphone market bootstrapped research and development in the automation of travel data collection that promises to automatically retrieve (i) accurate time-stamps and locations for journey origin and destination, (ii) Trip Path (route) with velocities along the journey and (iii) travel mode and trip purpose, \cite{berger_2015, zhao_2015, Prelipcean_2014, geurs_2015}. 
Additionally, to maximize uptake \cite{SEGMENT_006} (especially in the case of voluntary data collection), it is important to (i) avoid costs associated with data transfers; (ii) maintain anonymity and privacy; and (iii) consider battery energy consumption if personal mobile phones are being used.

In this paper we define and implement an algorithm, \Vjagg\footnote{``\vjagg'' is the Maltese word for ``journey''}, as an independent automated journey segmentation algorithm embedded as a mobility data collector, for both Android and iOS devices. \Vjagg is able to anonymously and seamlessly collect travel data from participants, using the device's GPS receiver and accelerometer. Mainly, it automatically segments the GPS trace into whole end-to-end journeys. Additionally, participants are given a summary of all journeys tracked and only those selected are uploaded to the server, guaranteeing full control on personal data. The algorithm is tested for accuracy in the journey detection task and for battery energy efficiency. The algorithm source code is released as open source under GNU GPL-v3.0 license\footnote{\url{https://github.com/michael-camilleri/vjagg}}.

Prelipcean et.\ al.\ \cite{PRELIPCEAN_2018} summarizes the available technology in automated travel diary collection, in terms of development, distribution and operations costs for both dedicated GPS receivers and smartphones. In the past, dedicated GPS receivers proved to be more cost effective than the use of smartphones, mainly due to the large development cost associated with the latter. However it is argued that open source solutions can lower these costs. MEILI \cite{Prelipcean_2018c} is probably the only open source system available right now. MEILI is a versatile research tool for setting up travel data surveys and computes the trip segmentation and classification on the server side. On the other hand the intention of \Vjagg is to support demand responsive transport (DRT) service providers, such as (with the customer's consent) quantifying missed opportunities.  The main focus in developing \Vjagg is therefore on accurate journey detection and on scaling up the collection of daily data for demand forecasting and extending services to new areas and corridors. \Vjagg  is therefore implemented and executed on the mobile device, such that data transfer is limited to journeys automatically suggested to and selected by the customer. Notwithstanding, we added trip purpose and mode manual functions such that \Vjagg (embedded as a mobility collector) can be used by transport researchers for travel behaviour studies.

The rest of the paper is organised as follows.
We first review related work in the area of automated and battery-aware journey segmentation algorithms (section \ref{sec_RELATED_WORK}) and then describe the problem of characterising journeys, Section \ref{S_PRELIMINARIES}.
This is followed by a detailed discussion of our battery-aware journey segmentation algorithm,  section \ref{S_ACTIVE_GPS}.
We then discuss our test setups and results (Section \ref{S_TESTING_AND_EVALUATION}), and finally give our conclusions and suggestions for future work.

\section{Related Work}\label{sec_RELATED_WORK}
We define \textit{journey} (or trip) as the path taken to travel from origin to destination, and a \emph{segment} as a part of a complete journey.
In this section we review methods that detect journeys and segments in a GPS trace as well as algorithms that minimise battery energy consumption.

\subsection{Detection of Journeys and Segments}
Intuitively, given a GPS trace, complete journeys are detected by identifying stops (i.e. GPS segments with zero speed). However stops do occur along journeys, (for example when queuing at junctions, or when changing mode) rendering the task a non trivial one.  Various methods have therefore been proposed. 
The first attempt in detecting journeys and trips solely from a GPS trace is described in 
\cite{Zheng_2008}. The authors define and make use of \textit{Change Point} segmentation, \ie where commuters change the transportation mode. Statistically it is shown that walking is engaged during most changes and therefore the algorithm detects walking (based on  velocity and acceleration thresholding) as the change point.
This method is further improved  using knowledge of the underlying transportation network \cite{Stenneth_2011}, and adding change of magnitude on heading and single travel-mode pattern-classifiers \cite{Zhang_2011}.
Lee et. al. \cite{Lee_2012} design a Variable-Rate-Localisation algorithm based on two key components: standstill detection, which uses a three-phase finite-state-machine based on GPS, accelerometer and Wi-Fi sensing, and an indoor/outdoor classification scheme.
In \cite{Biljecki_2013}, journeys are first extracted from the GPS trace, using features such as  signal shortage and long periods of idle time and then segmented into modes. 
All stops are considered as potential transition points and may result in a single-mode trajectory to be segmented into many shorter pieces, but consecutive segments of the same modes are eventually merged.  Similar experiments are reported in \cite{Rasmussen_2013, Nitsche_2014, Safi_2014}.
In most of the reported work, the parameters and thresholds are chosen from experience.
On the other hand, the thresholds are determined by a K-means algorithm in \cite{Stenneth_2012} and \cite{Xiao_2015} carry out a parameter search over a discrete grid to optimise the accuracy in detecting trip ends.
In general, the literature lacks a comparison of algorithms mainly due to the lack of a standard and suitable method to compare them \cite{Prelipcean_2016}.

\subsection{Battery-Aware Algorithms}

In this section we review the literature on battery-aware computing related to our work, i.e. in location tracking, where the power-hungry GPS sensing module is used.
In general, GPS battery-aware methods can be classified as either those that make use of two or more sensing modes \cite{POWER_003, POWER_006, POWER_009, Lee_2012, Prelipcean_2014}, or those that make use of past history or spatial-maps \cite{POWER_007, POWER_010, POWER_011, POWER_012}.

A single modal Location-Aware State Machine is used in \cite{POWER_003} to throttle the GPS sensor when the user appears stationary, with limited power-saving results, whereas in \cite{POWER_006, POWER_009, Lee_2012} the schemes employ the accelerometer to detect periods of motion that triggers or throttles the GPS.
Finally, in \cite{Prelipcean_2014} the authors utilise an `equidistance' tracking scheme, to predict velocity and dynamically adjust the rate at which GPS samples are taken, thereby saving power even while the GPS is in motion. Accelerometer readings are used to turn off the GPS when no motion is detected. Unfortunately, the authors do not quantify the savings due to their algorithm, focusing their contribution on the mode-detection instead.

From the survey it was clear that historical and spatial-map based methods are efficient in the use of battery consumption. However these schemes work well for coarse user-localisation, where the emphasis is on identifying where the user is at different periods of the day, often with respect to general key places and most of the time the detailed traces are compromised.  In our case we want to preserve accurate origin and destination points as well as a detailed and accurate trip trace.
Our algorithm is inspired from works that make use of various sensors, namely the GPS and accelerometer sensors. However whilst in the works reviewed above, the algorithms sample the accelerometer continuously, our algorithm makes use of one sensor at a time, \ie while journey tracking is active, it is the GPS itself that identifies periods of no motion and turns itself off.

\section{Problem Definition}
\label{S_PRELIMINARIES}
In this section we give a mathematical representation of the problem and our proposed solution and describe the issues considered in the design of the algorithm.

\subsection{Characterising Journeys}
\label{SS_CHARACTERISING_JOURNEYS}
\begin{figure}[!ht]
\centering
\begin{subfigure}[b]{0.6\textwidth}
	\includegraphics[width=\textwidth]{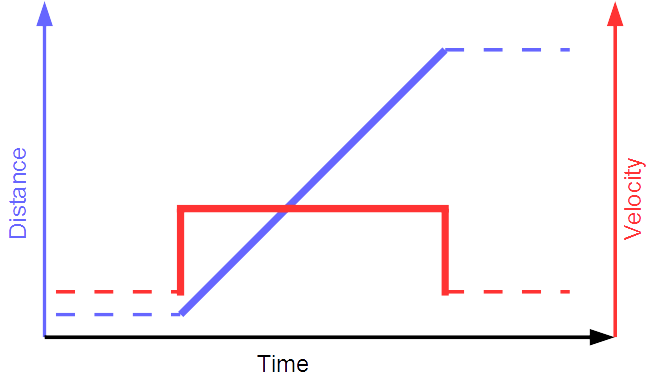}
	\caption{}
\end{subfigure}
\hfill
\begin{subfigure}[b]{0.3\textwidth}
	\includegraphics[width=\textwidth]{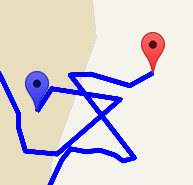}
	\caption{}
\end{subfigure}
\caption{(a) Idealised Journey Characterization, in terms of distance travelled (blue) and velocity profile (red). The solid component is what we would classify as a journey/segment. (b) Typical example of noise at the start/end of a journey.}
\label{FIG_JOUR_CHAR}
\end{figure}

The location sensor (in our case, the device's GPS receiver) returns a sequence $L=(l_i)_{i=1,2,\dots,N}$ of locations represented by the vector $l_i = (t_i,\phi_i,\lambda_i) : t_{i+1}>t_i$, denoting respectively the time-stamp, latitude and longitude coordinates. We will often summarise the spatial components of $l_i$ by $X_i=(\phi_i,\lambda_i)$.
We also define $L_a^b$ to be a subsequence of $L$ given by
\begin{equation}\label{EQ_ORDERED_SUBSEQUENCE} 
  L_a^b = (l_a,l_{a+1},\dots,l_b)
\end{equation}
with $1 \le a < b \le N$.
Additionally, we define the distance function $d(L_a^b)$ to be the sum of individual distances between each of the sub-sequence locations:
\begin{equation}
d(L_a^b) = \sum_{n=a}^{b-1} \hav\left( X_{n+1}, X_n \right)
\end{equation}
where $\hav(a,b)$ denotes the haversine distance operator between two locations $a$ and $b$ \cite{HAVERSINE_001}. We can now define a journey $J$ as a sub-sequence $L^{b}_{a}$ which satisfies the following two conditions:
\begin{gather}
d(J) \geq D \label{EQ_JOURNEY_DISTANCE} \\
\nexists\  (a^*, b^*) : a^* < a < b < b^*,\ d(J) \approx d(L_{a^*}^{b^*}) \label{EQ_SUBSET_CONDITION}
\end{gather}
The first condition, \eqref{EQ_JOURNEY_DISTANCE} enforces that within a time-frame there is `significant motion', represented by the threshold $D$. On the other hand, \eqref{EQ_SUBSET_CONDITION} ensures that no stationary periods at the ends of the journey contribute to the journey itself.
Figure \ref{FIG_JOUR_CHAR}(a) shows an idealised sample-set $L$ where the subject is initially idle, then moves for some distance before stopping. In this case, we wish to identify the `discontinuities' in the samples as the start and end of the journey. It follows that a journey is defined by those sections of the data where the velocity is above a threshold.\\
Mathematically, we can achieve this by numerically differentiating the location sample set $L$. We define the instantaneous speed of a point $l_i$ to be:
\begin{equation}
\left|\frac{\partial(X_i)}{\partial(t)}\right| \approx \speed{X_i, X_{i-1}} = \frac{\hav(X_i, X_{i-1})}{t_{i} - t_{i-1}} \quad \forall\ 1 < i \leq |L| \label{EQ_VELOCITY_ESTIMATE}
\end{equation}
Note that in some cases, we shorten the notation as $v_i = \speed{X_i, X_{i-1}}$ and  that the way we chose to approximate the derivative implies that we have one less velocity sample than position samples because $v_1$ is undefined. We can then rewrite our journey conditions, \eqref{EQ_JOURNEY_DISTANCE} and \ref{EQ_SUBSET_CONDITION}, as the contiguous sub-sequence of points where the speed is larger than our threshold:
\begin{equation}
J = \left\lbrace l_n \right\rbrace : v_n > V \quad \forall \ a \leq n \leq b \label{EQ_VELOCITY_THRESHOLD}
\end{equation}
where the symbols have the same meaning as before.

In implementing the above scheme, we have to deal with a number of practical issues.
First off, the GPS signal is characterised by numerous inaccuracies: we assume this to be Additive White Gaussian Noise (AWGN).
This is exacerbated by a `canyoning effect' in locations surrounded by high-rise buildings \cite{GPS_2011}.This phenomenon can lead to spurious jumps in locations, especially when the GPS signal is temporarily lost (see Fig. \ref{FIG_JOUR_CHAR}(b)).
This is particularly significant since the proposed speed-based thresholding and averaging may not easily mitigate this form of noise, due to the high-magnitude jumps.

In our discussion in Section \ref{SS_CHARACTERISING_JOURNEYS} we assumed that there is a definite starting and ending point of a journey, identified by clear changes in velocity. In practice this is rarely the case. Events such as stopping at junctions, bus-stops, pedestrian crossings, etc all trigger the end of a trip or segment and a simple velocity threshold is inadequate, yielding too many false-positives (in terms of end-triggers).
Additionally, our definition of $J$ in section \ref{SS_CHARACTERISING_JOURNEYS} applies more to a single leg of a journey, rather than a necessarily complete journey \cite{Prelipcean_2018c}.
Hence, we require methods for concatenating consecutive segments into single journeys.

\subsection{Algorithmic Overview}
\label{SS_ALGORITHMIC_IMPLEMENTATION}

The \Vjagg journey Identification Algorithm (VIA) provides the core implementation to address the goal of storing faithful GPS traces of whole journeys, whilst taking the following goals into consideration:
\begin{inparaenum}[(i)]
\item \label{ITEM_REDUCE_MEMORY} minimise device memory usage,
\item \label{ITEM_AVOID_INTENSIVE} avoid intensive computations during logging,
\item \label{ITEM_BATTERY_EFFICIENT} minimising battery energy consumption.
\end{inparaenum}

\begin{figure}
\centering
\includegraphics[width=0.8\textwidth]{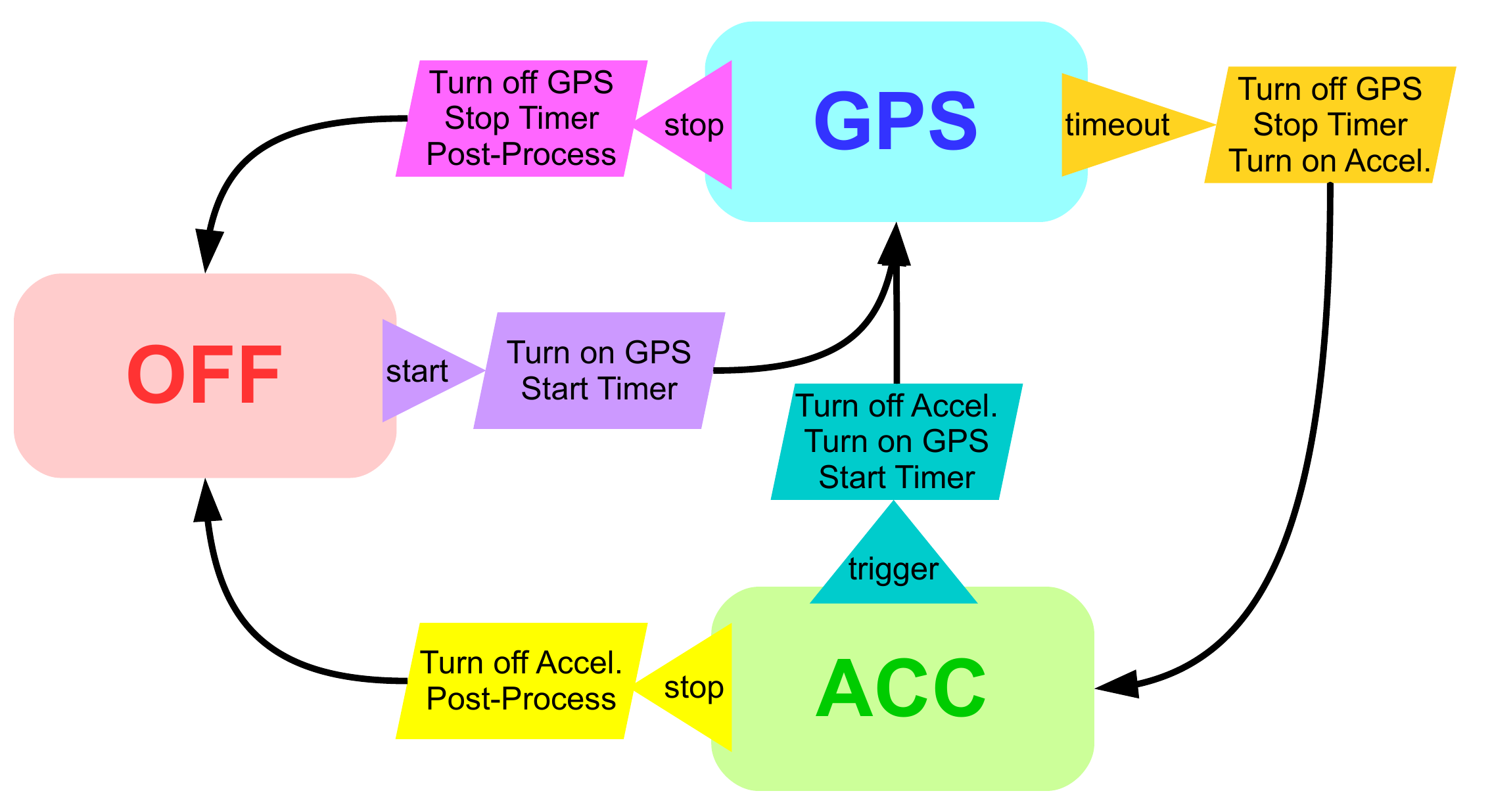}
\caption{Global State Transitions}
\label{FIG_STATE_MACHINE_GLOBAL}
\end{figure}

In order to achieve the above, we adopt a hierarchical approach, and implement a multi-level Finite-State-Machine (FSM), fig. \ref{FIG_STATE_MACHINE_GLOBAL}, making use of multiple sensing modalities. At the highest level, the FSM operates in one of three states:
\begin{inparaenum}[(i)]
\item \textbf{OFF} : The Tracking Service is off (default state). Once tracking is enabled (manually by the user or automatically via an alarm), we transition to \textbf{GPS}.
\item \textbf{GPS} : The Tracking Service is based solely on GPS sensing. In this mode, the algorithm is actively tracking journeys. The algorithm may exit this state if the user stops tracking or if it detects that the user is not travelling, in which case it switches to the \textbf{ACC} state.
\item \textbf{ACC} : The Tracking Service is solely using the Accelerometer (GPS is off), and looking for the presence of `significant motion' which potentially is the start of a journey. If motion is detected, the state goes back to \textbf{GPS}. These components are defined in the next section.
\end{inparaenum}

\section{Active Journey Tracking}
\label{S_ACTIVE_GPS}
The main component of VIA is the GPS-based tracking of journeys. 
In order to get the best balance between journey segmentation accuracy, data storage requirements and computational efficiency, the implementation follows a two-stage process.
\begin{enumerate}
\item An \emph{optimistic} online algorithm is employed during the active GPS sensing process.
This consists of a finite-state-machine which identifies periods of idle locations (no-motions), journey start/stop triggers and the periods in between.
\item The generated segments (i.e tentative journeys) are offloaded to temporary file storage, which are then post-processed using a multi-pass filtering scheme once the \textbf{GPS}-state is terminated.
\end{enumerate}
The emphasis of the online phase is to identify \textit{tentative} start and end of journeys, with a bias towards over-detecting journeys -- these can then be corrected using the offline post-processor which has a wider context.
The two-stage process also avoids the storage and post processing of the full GPS trace signal at once, including long entries of stationary points (which can be easily picked up by the online algorithm).
Note that the offline phase is still executed on the device!

\subsection{Algorithmic Motivations}

\subsubsection{Filtering out Noise in GPS data}
\label{SSS_DEALING_WITH_NOISE}
All the state changes of the on-line portion of the algorithm are governed (flow control) by the filtered (averaged) data samples $\hat{L}$, although we store all the raw GPS samples in $J$ for the detected journey. 
We choose to filter AWGN noise (see section \ref{SS_CHARACTERISING_JOURNEYS}) with an averaging down-sampler  (eqn. \ref{EQ_SAMPLING_AVERAGE}) , which worked well for the Markovian journey-start identifier (see section \ref{SSS_IDENTIFYING_START_TRIGGERS}).
\begin{equation}
\hat{X}_i = \frac{1}{W}\sum_{n=i+1-W}^i\left(X_n\right) \quad \forall i \in \left\lbrace Wk-1 \right\rbrace,\; 1 \leq k \leq \frac{N+1-W}{W} \label{EQ_SAMPLING_AVERAGE}
\end{equation}

In  equation (\ref{EQ_SAMPLING_AVERAGE}),  the average over a window size of $W$  samples is defined at every $W^{th}$ sample starting from $W-1$, $N$ is the sample size and $X_n$ is as defined before. The choice of $W$ is a trade-off between mitigating noise and guaranteeing a minimum data rate that is sufficient to identify start/end triggers.  

\subsubsection{Identifying Start Triggers}
\label{SSS_IDENTIFYING_START_TRIGGERS}
The start trigger is primarily identified with a velocity threshold defined in \eqref{EQ_VELOCITY_THRESHOLD}. Additionally, to mitigate the second form of noise (``canyoning effect'', section \ref{SS_CHARACTERISING_JOURNEYS}), we opt for a Markov-Chain decision process, whereby:
\begin{inparaenum}[(i)]
\item the velocity value must exceed the threshold for a number of successive windows (the MC state), and
\item the aggregate motion (the difference between the first and last data point in the MC state) must also exceed a threshold. 
\end{inparaenum}
The first condition seeks to reduce false-positives due to insignificant motions (for example shifting position while outdoors, resulting in significant velocity over a single or a few samples) while the second deals with the noise problem just mentioned (where  multiple high-velocities are present, but generally no significant motion).

Formally, for a set of $M$ consecutive samples from $\hat{L}$, a journey start is identified at down-sampled index $j$ iff these two conditions are met:
\begin{equation}
\begin{aligned}
v_n > V^i \quad \forall \; j \leq n < j+M \label{EQ_CUMULATIVE_VELOCITY_THRESHOLD} \\
\frac{\hav \left( X_{j+M-1}, X_j \right)}{t_{j+M-1} - t_j} > V^c
\end{aligned}
\end{equation}
where $V^i$ is the individual velocity threshold and $V^c$ the cumulative velocity threshold.

\subsubsection{Locating End Triggers}
Similarly, journey ending points are identified when the  velocity falls below a threshold and a hysteresis approach is employed to cater for traffic congestion and stops.
This entails that we do not search for an end-trigger until the \emph{displacement} within a number of samples $H$ (estimated as the displacement between the first and last point in the window) falls below a conservative threshold $D^H$. Once this happens we assume the journey segment has ended and attempt to locate a stop-trigger (at $j$) by running the Markov-Chain over the window $H$, as given by the conditions:
\begin{equation}
\begin{aligned}
v_n < V^i \quad \forall \; j < n \leq j+M \label{EQ_CUMULATIVE_VELOCITY_STOP} \\
\frac{\hav\left(X_{j+M}, X_{j+1}\right)}{t_{j+M} - t_{j+1}} < V^c
\end{aligned} .
\end{equation}
In general, the displacement windowing scheme works well, including when u-turns are present.  However the presence of stops when queuing at junctions and the passing through tunnels require further post-processing.
Due to our conceptual definition of a journey, another \textsl{sufficient} condition for identifying a journey as having ended is when the user enters a building (i.e. we ignore motion within buildings). This can be detected from the loss of the line-of-sight signal of GPS satellites.
VIA uses a time-out whereby if the GPS signal is lost for an extended period of time (to mitigate the effect of loss due to tunnels or high-rise buildings), the markovian-trigger-search scheme of (\ref{EQ_CUMULATIVE_VELOCITY_STOP}) is initiated.

\subsubsection{Concatenating Segments into Journeys}

In the process of concatenating segments into journeys we first consider what can lead to journey segmentation in the first place, and from there work our way towards concatenating them. The most obvious reasons, which are typically mode dependent, include:
\begin{inparaenum}[(i)]
\item \textit{On Foot} : taking a rest, meeting an acquaintance or waiting to cross the road.
\item \textit{Personal Vehicle}: idling in traffic congestion, waiting at junctions/stops/lights or dropping off/picking up
\item \textit{Public Transport} : idling in traffic congestion, waiting at junctions/stops/lights or regular scheduled stops.
\end{inparaenum}
A key realisation in all these instances above is that the sense of continuity is indicated by a short punctuation, both in time (period between stopping and restarting) and in space (distance between stopping and restarting point).
Our joining-algorithm is based on these heuristics.
Basically, two successive journey segments, $J_1$ and $J_2$ are deemed to be part of a larger whole \emph{iff}:
\begin{gather}
t^{J_2}_1 - t^{J_1}_{|J_1|} < T \\
\frac{K}{M}\sum_{m=1}^{M}v_{|J_1|-m}^{J_1} < \speed{J_{1,|J|}, J_{2,0}}
\end{gather}
where $T$ is a time-threshold, $K$ a distance threshold, and $M$ is the size (length) of an average.
The first condition enforces the time-constraint, while the second one ensures that it is realistic that given the speed of the user just prior to losing the signal, the second journey leg is a continuation of the previous one and the operation is recursive.

Finally, since we are not interested in very short commutes, we discard any journeys whose length is less than a threshold. In order to provide the best measure of journey length, we consider the journey bounds, indicated by a bottom-left ($bl$) and top-right ($tr$) corner pair rather than individual points:
\begin{equation}
\begin{aligned}
X^{bl} =  \left(\min_{n \in |J|} \left( x_n \right), \min_{n \in |J|} \left( y_n \right) \right) , \quad X^{tr} = \left(\max_{n \in |J|} \left( x_n \right), \max_{n \in |J|} \left( y_n \right) \right)
\end{aligned} ,
\end{equation}
where $x$ and $y$ are the cartesian co-ordinates of a location.

\subsection{Online GPS Logger}
The online portion of VIA is structured as a Finite-State Machine (FSM), triggered through GPS updates (handled by the underlying smartphone OS).
The finite-state formulation serves to synchronise the otherwise asynchronous callback mechanisms provided, while allowing efficient use of the single-threaded implementation. The VIA FSM  triggers on downsampled points, \ie state-changing decisions take place on the reconstructed $\hat{X}$ points rather than on the raw data at a base rate of 0.5Hz.
The emphasis in journey identification is on reducing False Negatives (at the expense of increased False Positives, which are handled by the off-line component).
The state of the FSM is governed by the following variables:
\begin{inparaenum}[(i)]
\item \textbf{Window Buffer}: $H$ holds the down-sampled point averages.
\item \textbf{Markov Chain}: $M$ keeps track of successive velocities within the $H$ Buffer
\item \textbf{Start\_ptr}: Keeps track of the sample (within $H$) at which a start of journey trigger was found.
\item \textbf{Journey}: $J$ conceptualises a journey segment, storing it to file as required.
\end{inparaenum}
The FSM itself consists of four states. In all states, the GPS is turned \textit{ON} and points are buffered, however not necessarily logged to file. The states are (in order of typical flow):
\begin{inparaenum}[(i)]
\item \textbf{Idle} : The $M$ Buffer is still filling up. This state is required because the Markov Chain logic requires at least two-samples to commence.
\item \textbf{Searching} : The Markov Chain is searching for a start trigger.
\item \textbf{Found} : The Markov Chain has found a start trigger, but not enough time-points have been retrieved to allow determination of end-triggers ($H$ buffer is filling up).
\item \textbf{Logging} : The journey is being logged to file.
\end{inparaenum}

\begin{figure}[!ht]
\centering
\includegraphics[width=0.9\textwidth]{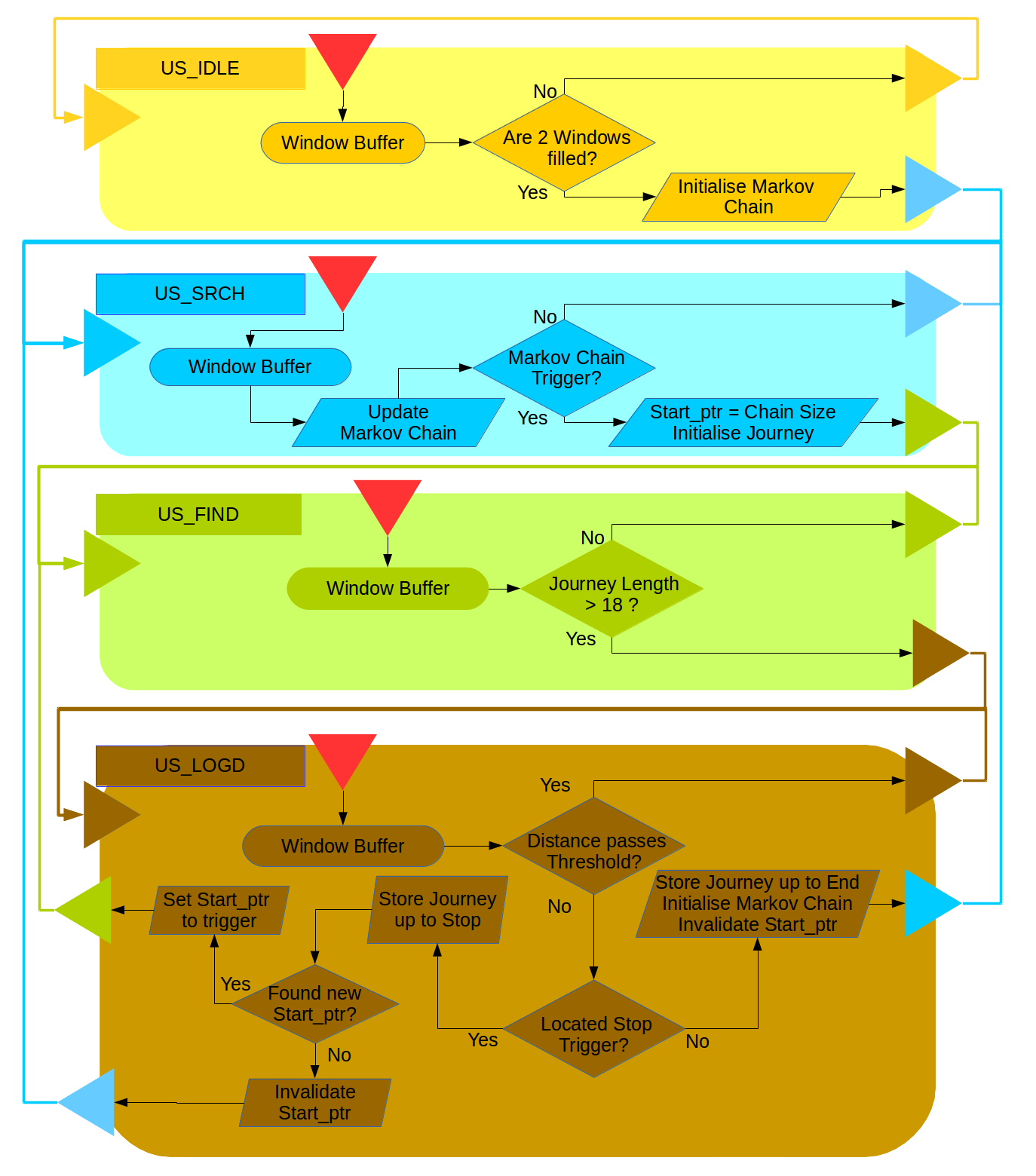}
\caption{Flowchart of the GPS Finite State Machine showing state progression under normal circumstances.}
\label{FIG_STATE_MACHINE_NORM}
\end{figure}

Figure \ref{FIG_STATE_MACHINE_NORM} depicts the flow of control between states.
Each state is represented by its colour, such that the colour of the operation itself serves to identify the state with which it is associated (including the transition to the next state). All operations with the same colour and within the same background signify completion within the same down-sample callback.
Transitions take place on any down-sample callback cycle (but not within the same callback) and are indicated by the coloured arrow-lines extending outside the state border.
Each state has an entry point (of its own colour) and a set of exit points (with the colour of the state that the machine will transition to on the next cycle).
Furthermore, the entry point for each state is not connected to the first operation to be performed while in that state: instead, this comes from the red-coloured entry point (down-sample callback), to emphasize that the state change happens on a callback boundary.

The State Machine starts in the \textbf{Idle} state (\textbf{\textsl{US\_IDLE}}), with the $H/M$ buffers flushed. 
Once the buffer is full, the machine transitions to the \textbf{Searching} state. The Markov Chain $M$ is initialised with the velocity between the first two windows.
Once the trigger is found, we set the \textbf{\textsl{Start\_ptr}} to the beginning of the appropriate window and transition to the \textbf{Found} state.
The \textbf{Found} state (\textbf{\textsl{US\_FIND}}) serves to buffer samples until there are enough for the end-trigger calculation.
Once a journey reaches the $H$ length, VIA transitions to the \textbf{Logging} state.
Finally, in the \textbf{Logging} state (\textbf{\textsl{US\_LOGD}}), VIA buffers the samples while checking for the distance threshold over a period of $H$ windows. If the distance threshold is met, then we stay in the same state. Otherwise, we test for an end of  journey. We run the Markov Chain $M'$ in reverse (starting from the oldest window and moving forwards in time) until we locate a stop trigger or we reach the present location. If no stop trigger is found, we simply store the journey up till the latest point and transition to the \textbf{Searching} state. Otherwise, we attempt to find another start trigger, since a new journey could have started in the mean-time. If one is found, we set the \textbf{\textsl{Start\_ptr}} and move to the \textbf{Found} state. Otherwise, we retain the Markov Chain state (with a potential partial trigger) and switch to the \textbf{Searching} state.

In addition, due to time-outs and the stop-event, a further asynchrony is introduced. The asynchrony refers to the fact that although the actual transition does happen when the down-sampler is idle (i.e. it does not interrupt an in-state operation as illustrated by a filled background in Fig.\ref{FIG_STATE_MACHINE_NORM}), it can happen any time in between calls (it is state independent). In fact, more often than not, it happens between successive GPS updates, implying only a partial down-sample (which must be explicitly catered for, since the algorithm runs at the down-sampled rate).

Besides indicating a potential end of journey, long gaps between GPS fixes could pose a problem for thresholds. The location callback itself is triggered only when there is a fix, and hence, if not called, will halt the FSM.
A watchdog is thus employed to identify when the signal is lost for an extended period of time. If the watchdog triggers, it checks whether the currently active state indicated a journey was being logged (i.e. we were in state \textbf{Logging}). If this is the case, then we attempt to find a stop trigger within our buffer and store the journey up to either the location of the stop trigger (if one is found) or to the end of the buffer. If no journey was active we flush the buffer (and transition to the \textbf{Idle} state).

\subsection{The Offline Post-Processor}
\label{SS_OFFLINE_LOGGER}

The Post-Processing algorithm is triggered when the user presses the Stop Tracking Button (after the on-line algorithm terminates) or the \textbf{ACC} trigger kicks in. The first task is to load all stored journeys from the temporary file generated by the FSM. The algorithm then executes a number of distinct routines:

\noindent\textbf{Threshold based on Distance (low):}
Initially, all journeys whose length is less than 50m are discarded. The low 50m threshold ensures that if a valid journey consists of multiple small parts (perhaps due to being stuck in traffic), it does not get eliminated in this first step.

\noindent\textbf{Journey Concatenation:}
The main aim of post-processing is to allow individual journey segments to potentially be joined into a single journey. Starting from the next to last journey and moving backwards, if the duration between the two journeys is less than a threshold and the distance between the two journeys is such that the average velocity at the end of the first journey (up to a tolerance factor, currently set at 120\%) indicates that the starting point of the second one is a viable continuation, then the journeys are concatenated.

\noindent\textbf{Threshold based on Distance (high, 500m):}
Another threshold on distance is performed.
This together with the initial thresholding operation, provides a hysteresis threshold.

\noindent\textbf{End Trimming:}
Finally, in order to mitigate the spurious jumps which occur while the GPS system is achieving a stable fix, the ends are trimmed for points with velocities in excess of 20m/s, up to a limit of three eliminations (to prevent eliminating the entire journey).

\subsection{Battery-Aware Algorithm}
\label{S_BATTERY_AWARE_TRACKING}
In this section we describe the battery energy consumption savings features of the VIA. The algorithm depends on the availability of a sensory input, which yields a distinct output value when the device is idle (motionless) and when it is in motion (travelling). We use the accelerometer, which is found in virtually all smartphones ($> 99\%$\footnote{https://opensignal.com/sensors/library/accelerometer}).

From an information perspective, the GPS is redundant when the user is stationary and when the GPS cannot obtain a reliable fix, such as when the user is indoors, the latter being dealt with in the context of the journey detection portion of the algorithm. We deal with the former by means of another watchdog time-out condition, which triggers when no journey has been active (VIA state is US\_IDLE or US\_SRCH) for an extended period of time (in our case five minutes). When this happens, the GPS is switched off, and the algorithm transitions to looking for significant motion by way of the accelerometer.

The problem of \textit{identifying significant motion} is difficult due to (a) the presence of noise (which is especially pronounced in the accelerometer), \textsl{and} (b) the term stationary may not necessarily mean perfectly motionless. In short we wish that no motion is signalled when the device (a) is on a table, (b) is in the user's bag/garment pocket who is sitting or standing, (c) vibrates while in the user's pocket, \textsl{or} (d) is briefly checked by the user. In particular handling conditions (c) and (d) reduces false-positives and the GPS turning on unnecessarily. Conversely, we wish to detect motion when the user walks or drives with device in hand, bag or garment pockets.  In VIA the emphasis is on reducing false negatives, since we seek to pick up the starting point of a journey as accurately as possible.

\begin{figure}[!h]
\centering
\includegraphics[width=0.8\textwidth]{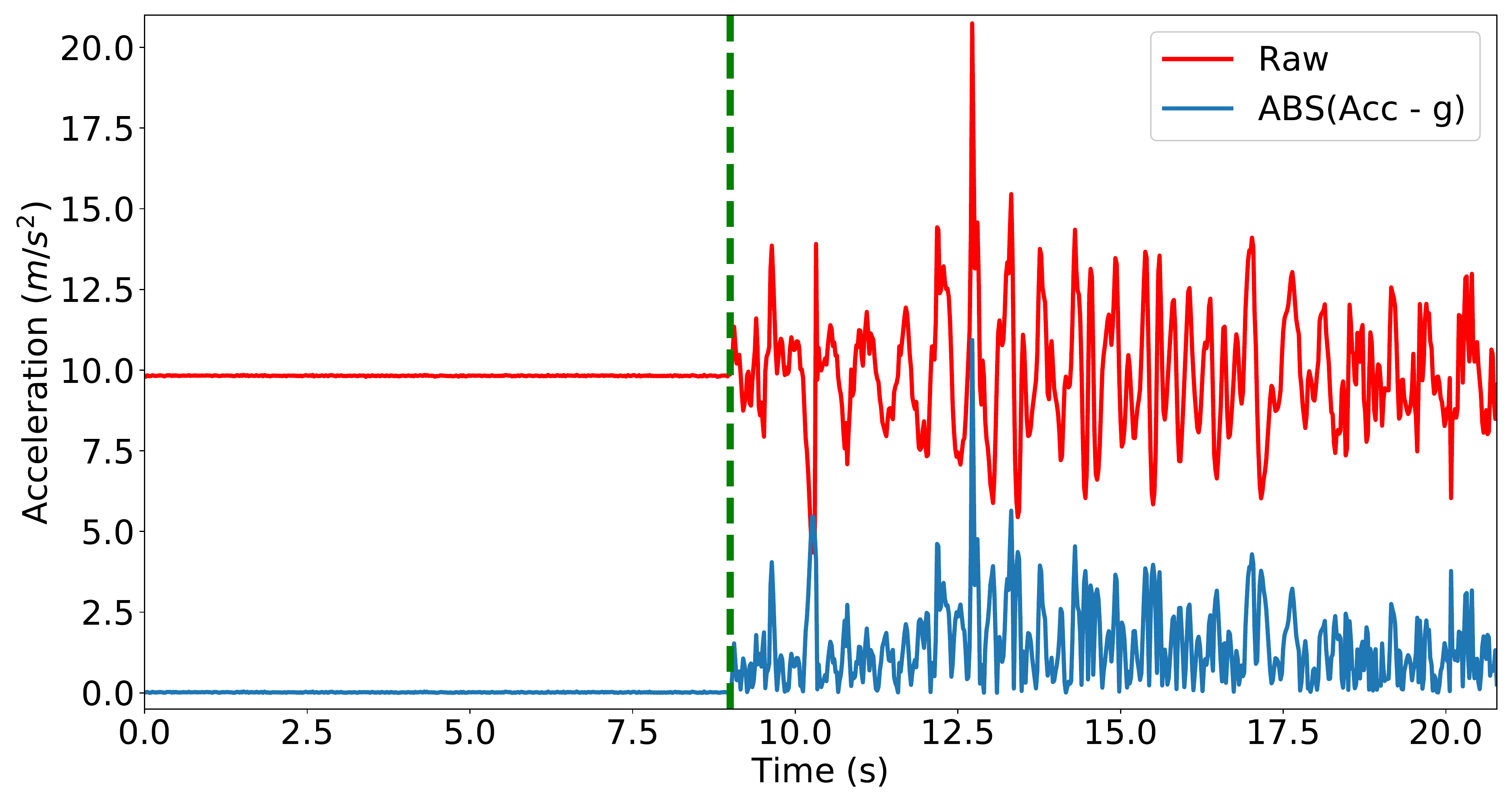}
\caption{Typical Acceleration Profile: the dotted line indicates the transition from standstill to motion. The raw acceleration is shown in red, with blue being the filtered version (subtracting $g = 9.81$ and taking absolute value).}
\label{FIG_ACCEL_PROFILE}
\end{figure}
\begin{figure}[!hb]
	\centering
    \begin{subfigure}[b]{0.48\textwidth}
    \includegraphics[width=\textwidth]{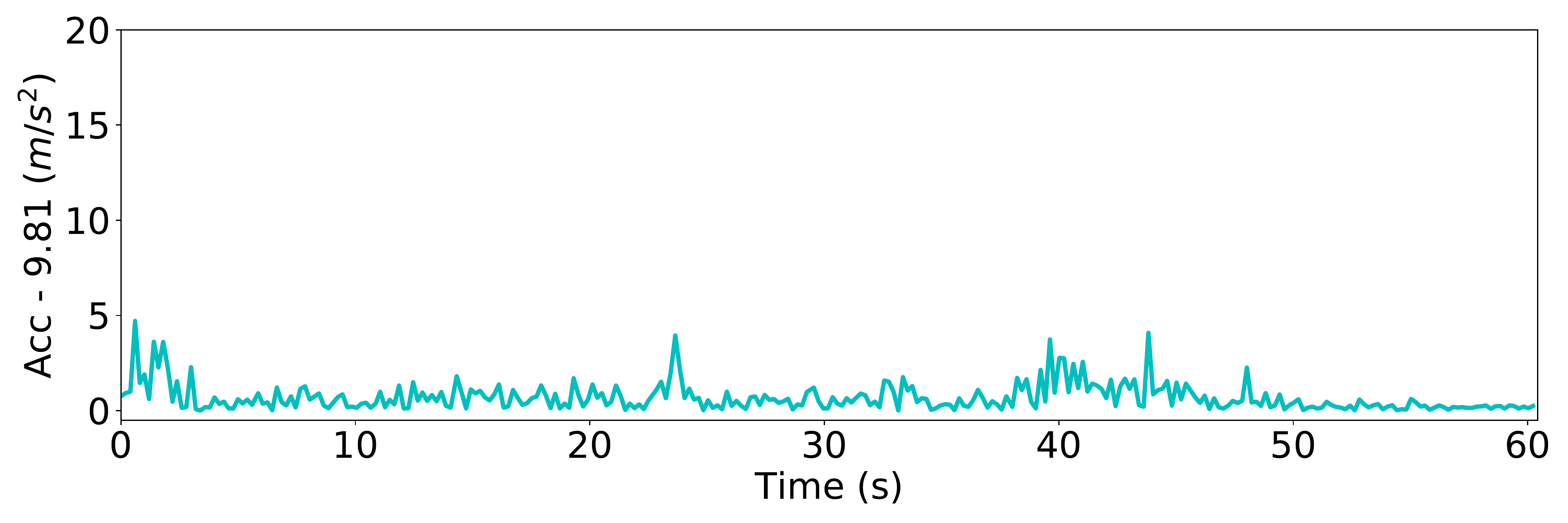}
    \caption{Travelling by Bus}
    \end{subfigure} \hfill %
    \begin{subfigure}[b]{0.48\textwidth}
    \includegraphics[width=\textwidth]{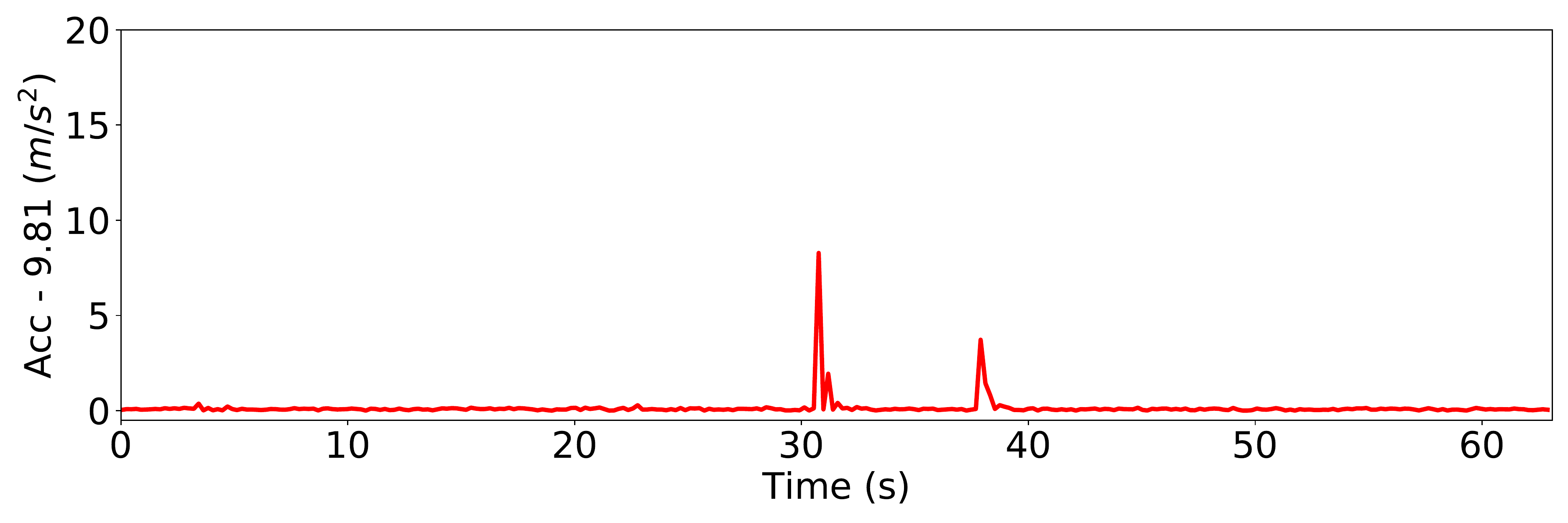}
    \caption{Device on Table}
    \end{subfigure}
    \begin{subfigure}[b]{0.48\textwidth}
    \includegraphics[width=\textwidth]{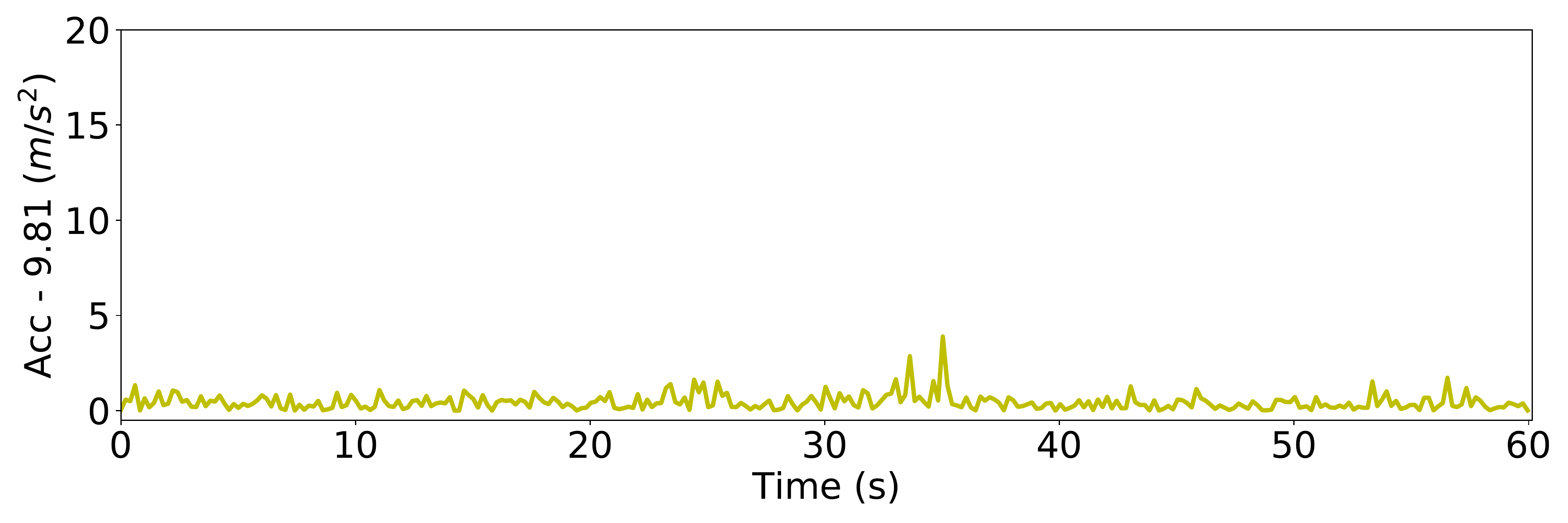}
    \caption{Travelling by Car}
    \end{subfigure} \hfill %
    \begin{subfigure}[b]{0.48\textwidth}
    \includegraphics[width=\textwidth]{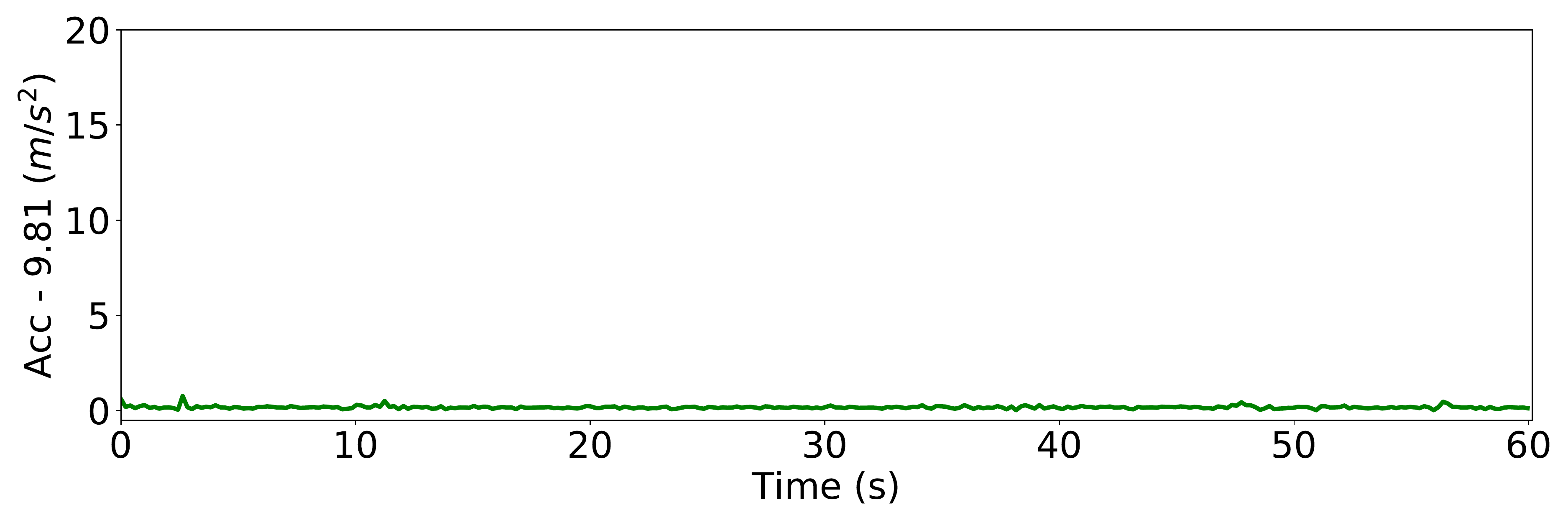}
    \caption{Sitting at Work}
    \end{subfigure}
    \begin{subfigure}[b]{0.48\textwidth}
    \includegraphics[width=\textwidth]{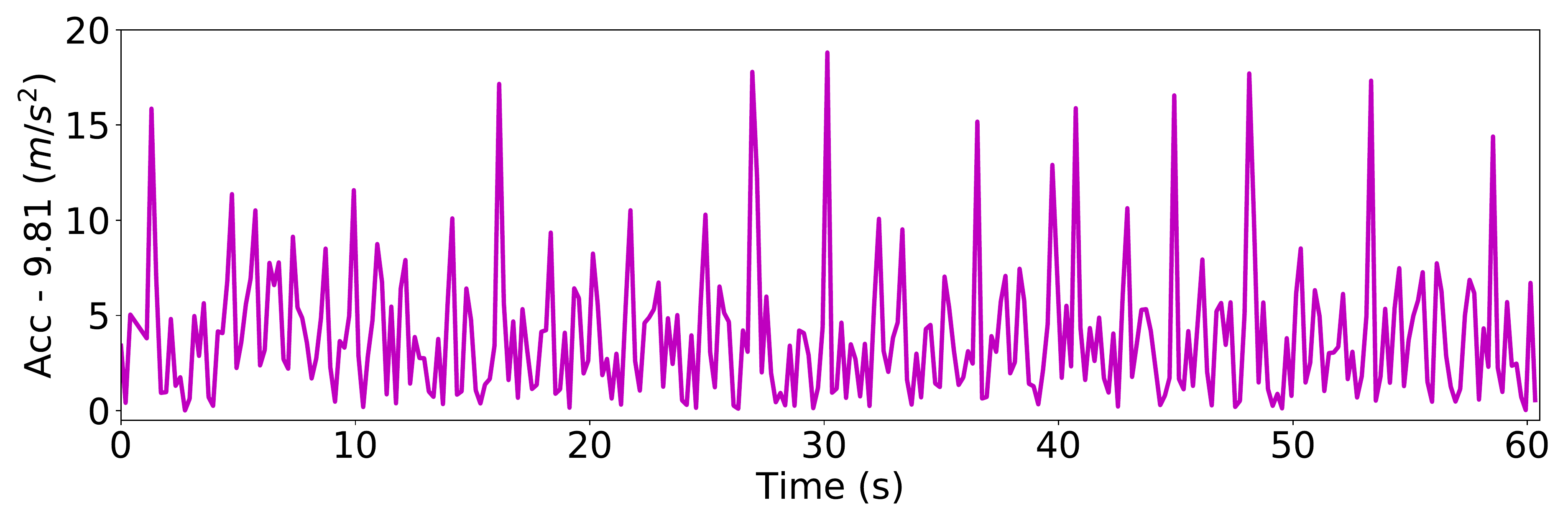}
    \caption{Travelling on Foot}
    \end{subfigure} \hfill %
    \begin{subfigure}[b]{0.48\textwidth}
    \includegraphics[width=\textwidth]{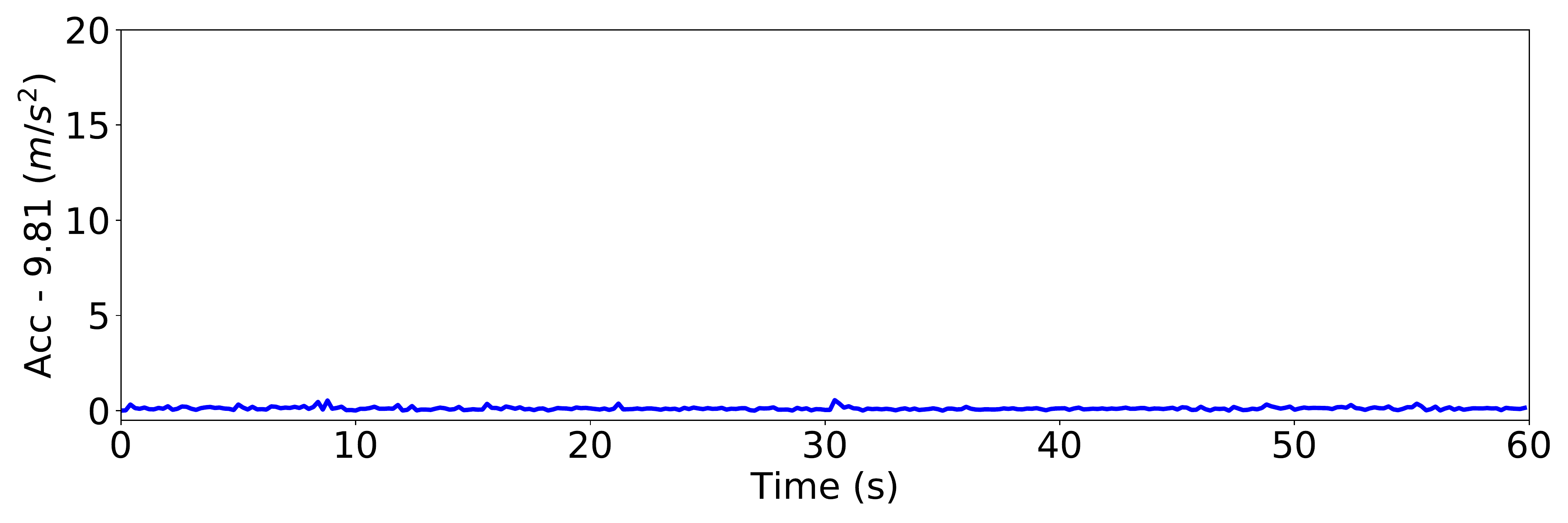}
    \caption{Sitting/Standing at meeting}
    \end{subfigure}
    \caption{Acceleration profiles for Moving (Left) and Motion-Less (Right) scenarios.}
    \label{FIG_ACC_MOTION}
\end{figure}
\begin{figure}[!ht]
\centering
\includegraphics[width=0.9\textwidth]{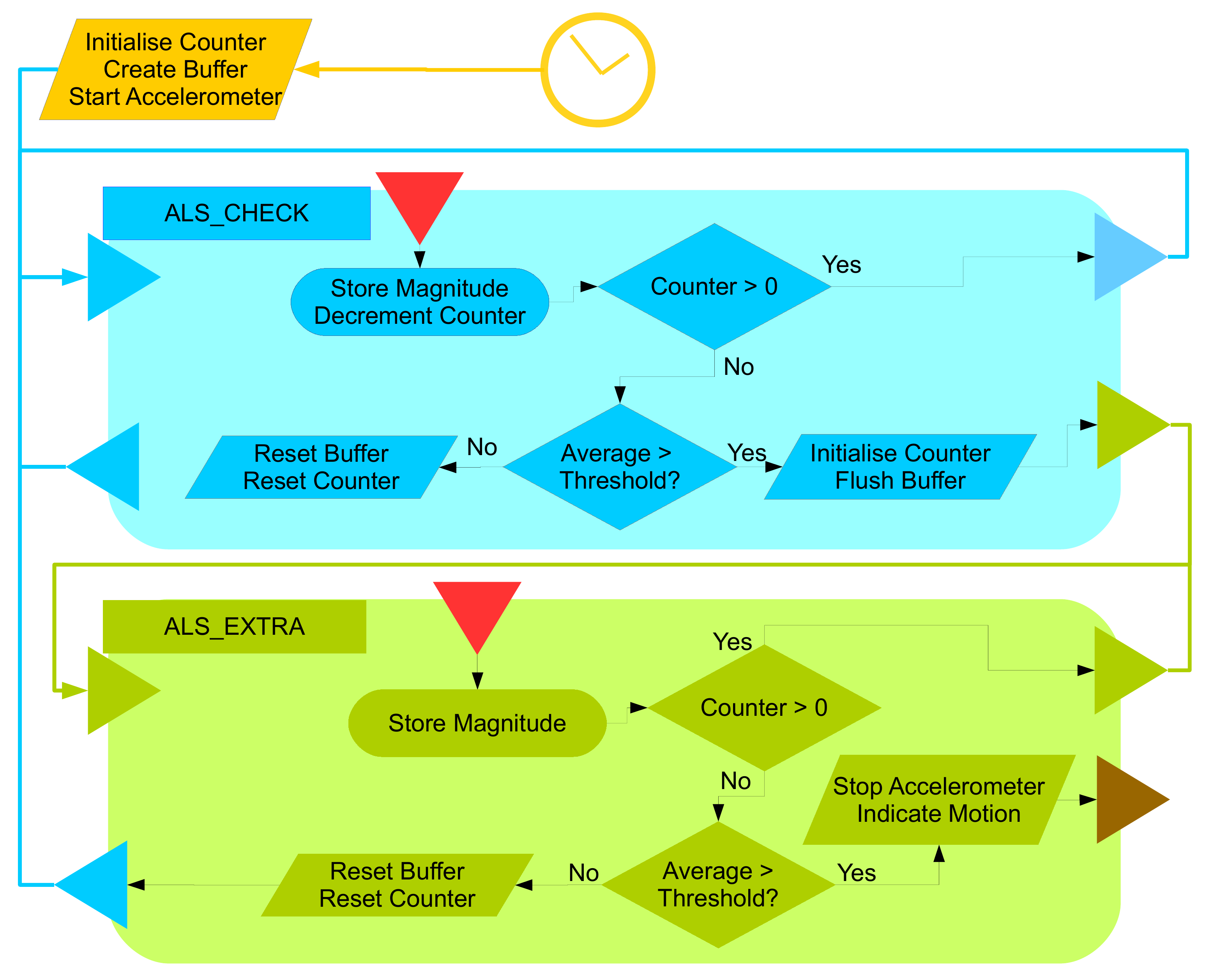}
\caption{Detection of Significant Motion (trigger from Accelerometer to GPS). (The colour codes here are independent of those in Fig. \ref{FIG_STATE_MACHINE_NORM})}
\label{FIG_STATE_MACHINE_TO_ACTIVE}
\end{figure}

Fig.\ \ref{FIG_ACCEL_PROFILE} displays a typical acceleration-magnitude profile (blue trace) of a Samsung Tablet, transitioning between static to hand-held, where we note that while the raw average appears to remain the same, the average deviation from the nominal 9.81$m/s^2$ is detectable.
Fig.\ \ref{FIG_ACC_MOTION} gives acceleration profiles in a number of scenarios, while the decision logic is given in Fig. \ref{FIG_STATE_MACHINE_TO_ACTIVE}.
Optimisation of thresholds/parameters was achieved through simulated annealing and random search (described further below).
The Finite-State-Machine (FSM) goes through three distinct states.
\begin{inparaenum}[(i)]
\item \textbf{{INIT}} : This is an intermediary initialisation stage which is executed when the GPS is turned off.
\item \textbf{{CHECK}} : This state provides a low threshold for early detection of motion (based on a Markovian threshold) which can then later be verified in the \textbf{EXTRA} state.
\item \textbf{{EXTRA}} : This state consists of an extended sample run, using a different threshold to confirm or reject the original hypothesis arrived at in state \textbf{CHECK}.
\end{inparaenum}
The decision to use a two-stage checking scheme follows from the intuition that even in motion-less scenarios, there may be occasional spikes, albeit of short duration (see Fig.\ref{FIG_ACC_MOTION}(f), when compared to (c)) which can throw off a single threshold. This also spurred the option to test with multiple successive averages in the \textbf{EXTRA} state.

\section{Tuning and Evaluation}
\label{S_TESTING_AND_EVALUATION}
In this section we discuss the tuning process of the various free parameters and their experimental validation, and describe and evaluate the empirical  experiments carried out to determine the efficacy of  VIA on real-world data.

\subsection{Experimental setup}
Our tests were carried out on a range of devices from different manufacturers and running on various OS versions, as noted in Table \ref{TAB_LIST_OF_DEVICES}.
We collected data from three main modalities, namely the raw sensors, battery-usage and pen-and-paper diaries for validation.

\begin{table}
\begin{center}
\begin{small}
\caption{List of Android Test Devices. Specifications retrieved from \textit{www.gsmarena.com}}
\begin{tabular}{|l|c|c|c|c|}
\hline
\rule{0pt}{3ex}\textbf{Device} & \textbf{Tab E 9.6"} & \textbf{Tab A 10.1"} & \textbf{Pulp 4G} & \textbf{J5} \\
\hline
\hline
\rule{0pt}{3ex} \textbf{\textsl{Device Code}} & T1 & T2 & S1 & S2 \\
\hline
\rule{0pt}{3ex} \textbf{\textsl{Quantity}}    &  1 &  1 &  1 &  3 \\
\hline
\rule{0pt}{3ex} \textbf{\textsl{Manufacturer}} & Samsung & Samsung & Wiko & Samsung \\
\hline
\rule{0pt}{3ex} \textbf{\textsl{Model}} & T561	& T585 & -- & J510FN \\
\hline
\rule{0pt}{3ex} \textbf{\textsl{OS Version}} & 4.4.4 & 6.0.1 & 5.1.1 & 6.0.1 \\
\hline
\rule{0pt}{3ex} \textbf{\textsl{CPU Cores}} & 4 & 8 & 4 & 4 \\
\hline
\rule{0pt}{3ex} \textbf{\textsl{CPU Speed}} & 1.3GHz & 1.4/1.0GHz & 1.2GHz & 1.2GHz \\
\hline
\rule{0pt}{3ex} \textbf{\textsl{Battery (mAh)}} & Li-Ion 5000 & Li-Ion 7300 & Li-Po 2500 & Li-Ion 3100 \\
\hline
\end{tabular}
\label{TAB_LIST_OF_DEVICES}
\end{small}
\end{center}
\end{table}

\noindent\textbf{Sensor Data:}
Verification of the algorithmic operation required multiple runs with different parameters on the data-sets. To this end, we incorporated an option in our wrapper application to save all sensor data (GPS fixes, satellite and accelerometer readings to file storage for later retrieval.
All readings were time-stamped, and offloaded to an external text file, which we then retrieve directly through the phone's file system.
In the interest of efficiency and file-size considerations, we cap the sampling rate at 0.5Hz for the Location and 5Hz for the acceleration, as indicated in literature, e.g. \cite{Prelipcean_2014}.

\noindent\textbf{Diaries:}
In order to fine-tune the parameters and validate the algorithm we required the addition of annotated journey data with clear starting and stopping points, kept by the annotators. To aid this, we also employ a `ping' feature within the `debug' application to allow the annotator to explicitly indicate the start or stop of a journey.

\noindent\textbf{Battery Consumption:}
Typically, power data is collected using elaborate custom-made hardware, e.g. \cite{POWER_001}, however, given the constraints of our project, we opted for a software based approach.
In this case, we collected battery percentage and voltage levels, and for all but T1 the instantaneous/average current drawn. Data was sampled at either 15 (IDLE-battery tests) or 1 (for all other tests) minute intervals using the \textsl{AlarmManager}.

\subsection{GPS Noise}
Although it was not our intention to fully characterise the noise process within the Location framework, we nonetheless ran a number of tests to determine whether the amount of noise would be significant in our algorithm. We ran two types of tests, in a static and dynamic setting.

\begin{figure}[!ht]
\centering
\includegraphics[width=0.9\textwidth]{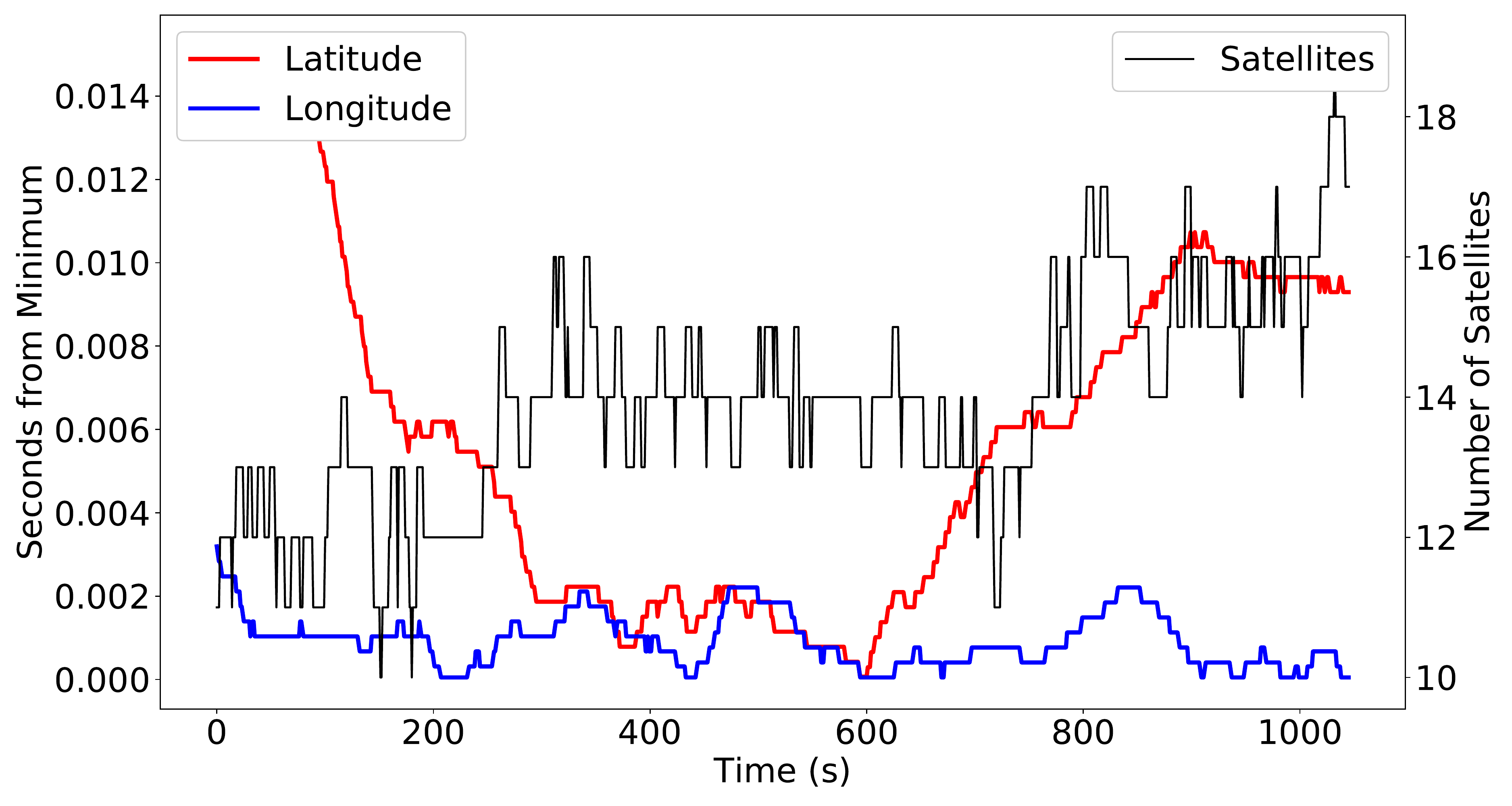}
\caption{Typical example of static noise for Device S2 under extreme foliage.}
\label{FIG_STATIC_NOISE}
\end{figure}

\noindent\textbf{Static Noise:} To quantify the  magnitude of noise in different static scenarios, we left the device in a position with a view of the sky (namely in direct view, partially obstructed by a building and under foliage) and computed the difference from a mean recorded location for a period of 10 minutes. This setup is based on the assumption that the noise is Gaussian distributed around the actual true value.
Results showed that, apart from a certain amount of drift with time in the recorded locations, there were no significant errors beyond $10^{-4}$ in latitude and $1.4\times10^{-4}$ in longitude. In fact, the satellite count as recorded by the framework was barely affected in each of the tested scenarios, varying between 11 and 16 satellites. Figure \ref{FIG_STATIC_NOISE} displays a sample run.

\begin{figure}[!ht]
\centering
\includegraphics[width=0.9\textwidth]{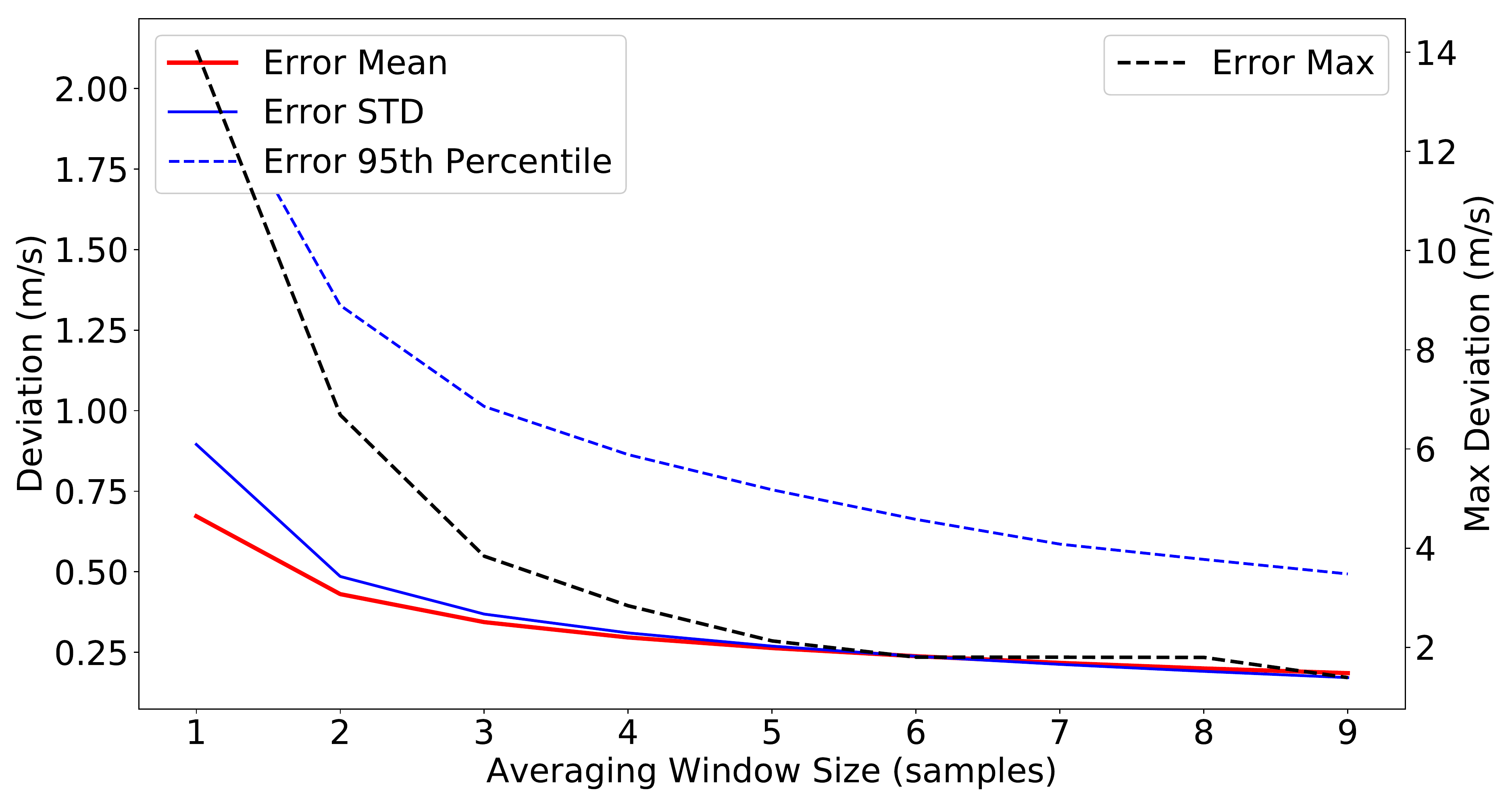}
\caption{Evolution of deviation from Nominal Velocity due to noise for varying average-window sizes. Note that \textit{max error} is on a different scale (right).}
\label{FIG_DYNAMIC_NOISE}
\end{figure}

\noindent\textbf{Dynamic Noise:} Potentially, in the VIA algorithm, the greatest detriment of noise would be in causing false-positive triggers for the start/stop detection algorithm. We therefore sought to characterise the effect of the noise on velocity computations from successive positions.
It was also desirable to test the system under severe canyoning effects which provide the worst case scenario for GPS tracking.  The tests where carried out in Valletta, characterised by a Manhattan Grid street pattern, narrow streets and relatively high buildings.
The experiments involved walking at a moderate constant pace alongside a block. Multiple circumnavigations of more than one block at a time were carried out, with the annotator pinging  at each corner. The experiments were repeated around different blocks.
The data, collected at intervals of 1s, was then filtered as follows. 
First, the straight-line motions between corners of the blocks, were separated into individual runs $R_i$.
For each of these runs, the mean velocity $V_i$ was calculated by dividing the straight-line distance between the corners (obtained online from Google Maps) by the total travel time (obtained from the pings registered by the researcher).
The individual points are then averaged over windows of varying sizes $w \in \{1...10\}$ to yield a down-sampled run $D_{i,w}$.
In order to increase the number of samples, for reliability of the computation, in windowing schemes, multiple runs $D^k$ were computed started at successive points.
Finally for each pair of points $d^{j,j+1}_{i,w}$ within $D_{i,w}$, the first-order approximation of the velocity was computed, and the deviation from the nominal velocity $V_i$ recorded.
In summary, the mean discrepancy for a particular window size $w$ was computed as follows:
\begin{equation}
\tilde{V}_w = \frac{1}{|W||I|}\sum_{k=0}^{w-1}\left\lbrace\sum_{i=1}^{|I|}\left(\frac{1}{|D_{i,w}^k|}\sum_{j=1}^{|D_{i,w}^k|-1} \left| \frac{d^{k,j+1}_{i,w} - d^{k,j}_{i,w}}{t^{k,j+1}_{i,w} - t^{k,j}_{i,w}} - V_i \right| \right)\right\rbrace
\end{equation}
Figure \ref{FIG_DYNAMIC_NOISE} illustrates the effect of window-size on the estimated velocity `\emph{error}'.
As can be seen (and mostly from the maximum error), a window size of 3 should provide adequate filtering (with the 95$^{th}$ percentile lying within 1.1 m/s from the nominal) while not delaying triggers excessively (due to too large windows).

\subsection{Identifying Journey Starts}
The parameters of interest here are the three related to the aforementioned Start Markov Chain: i.e. the number of velocity points to consider, the instantaneous velocity threshold and the total velocity threshold.
The thresholds here are set for typical walking speeds: given the average speed of 1.3m/s, we cap both the instantaneous and total velocity at 1m/s, which must both be exceeded to start recording data.
This is a conservative threshold, but we chose it because we prefer to generate False Positives (\ie starting tracking when no journey actually exists), which can be handled by the offline post-processor, rather than False Negatives (which would miss out on a journey). 
The choice of three time-steps was based on visual inspection of the GPS traces during start/stops of journeys as well as typical motion patterns for individuals in open spaces. Specifically, given the 0.5Hz sample rate and the further down-sampling by 3, the length of the chain corresponds to a time-period of 24 seconds which is adequate to filter out short motions but can capture intent to start a journey.

\subsection{Determining Journey Stops}
In the stop-detection, preference is given to False Negatives (\ie not detecting a stop) rather than False Positives.
With regard to the detection window, we employ 2.5 minutes worth of data (25 down-sampled points), and a threshold distance (i.e. distance between first and last point in the set) of 30m.
The motivation behind this scheme is targeted mostly at vehicular travel: specifically, it seeks to avoid detecting slow traffic or stopping at junctions/lights as journey termination.
The values were intuitively set based on traces of journey data.
Given this, we then retroactively seek to find the actual point of journey termination, using the same Markovian scheme as for journey starts, although with the conditions reversed.
In this case it was found adequate to use the same thresholds as above: \ie 3 velocity windows, with instantaneous and total velocity thresholds set at 1m/s.

\subsection{Satellite Indications}
The number of satellites in line-of-sight are a key indication of journey termination (due to the user having entered a building), but at the same time, can generate False Negatives (such as tunnels).
Theoretically a GPS fix can be achieved with a minimum of 4 satellites: however, this threshold also has to consider the difference between typical indoor and outdoor receiving status.
In tests carried out it emerged that in good view of the sky, the satellite count could be as high as 11 or 12 satellites, while indoors this falls to 0 or sometimes 1.
Hence, we decided to cut-off at a value of less than 5.

The determination of the time-out at which to signal such a journey end has to do with the tunnel problem. This is hard to quantify, as it depends not only on tunnel lengths but also the vehicle speed and the presence of traffic.
We employed a time-out of 40 seconds which worked well for our use-case.
At the same time, we choose to mitigate the problem using the off-line post processor which is able to join together journey segments.

\subsection{Concatenating Journeys}

The journey concatenation scheme is mainly designed to address the problem of vehicular journeys being divided into smaller segments, due to temporary signal loss.
From the data, we observed that the velocity magnitudes at the splitting points are similar.
This condition forms the basis for our concatenation algorithm.

More specifically, concatenation depends on two parameters; (i) the time difference between successive journey segments, manually set to 2 minutes, and (ii) the end/start velocity ratio, set to 1.2.
It should be noted that this scheme does not handle the case where the segmentation is due to being stuck in traffic or at junctions.
This is because, in this case the conditions are typically the opposite (short wait times, and excessive velocity differences).
Instead, this is designed to be catered for by the stopping hysteresis in the on-line algorithm.

\subsection{Filtering out irrelevant journey segments}
Very short journeys (e.g. moving between buildings on a small campus) are typically non-informative for the scope of our use case and we delete journeys that are less than 500 metres.
We also  trim portions of journeys that are due to spurious jumps in GPS traces, that occur due to the location service making use of both GPS and other less accurate sensors (such as Wi-Fi access-point information or mobile cells), and typically characterised by a rapid jump, with velocities in excess of 20 m/s. While this velocity is itself perfectly normal for vehicular travel, these typically happen towards the beginning or the end of a trip (when there is GPS signal loss), and where typically, velocities are still low mostly due to the first mode of transport, i.e. walking. Hence, we discard such points at the beginning (first three points) and end of the journey (last three points) whose velocity exceeds this threshold.

\subsection{Battery Consumption Profiling}
The battery-aware sub-system of our base Journey Segmentation algorithm comes at a price in data accuracy. Extensive testing was carried out to find the best trade-off between battery efficiency (which would impact on user uptake) and accuracy of data.

\subsubsection{Idle Battery Consumption}
\begin{figure}[!ht]
	\centering
    \includegraphics[width=\textwidth]{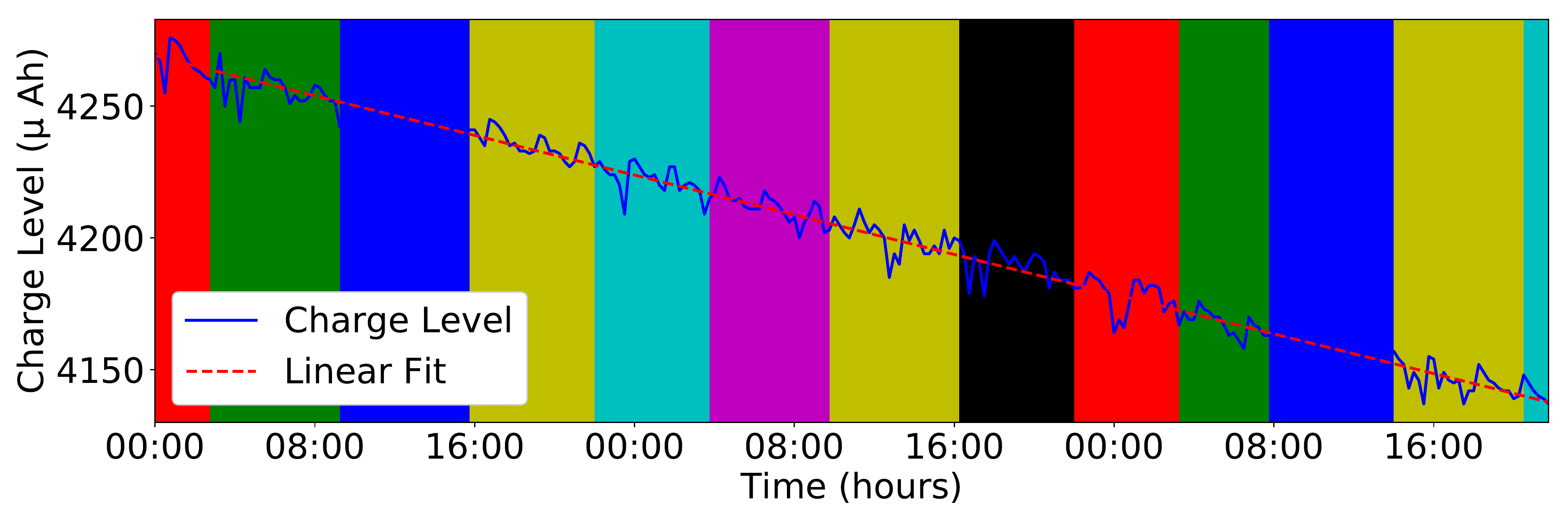}
    \caption{Idle Discharge for T1. The coloured bars indicate different battery-\% readings. Note how due to their inconsistent width, they are not directly indicative of consumption.}
    \label{FIG_IDLE_DISCHARGE_V}
\end{figure}

We ran tests with the device in standby to identify the power-consumption of the device under idle conditions (no applications running, Wi-Fi/3G off, GPS switched on but not polling).
We also allowed the OS to handle all sleep control, including DOZE\footnote{Note that our algorithm actually disables DOZE, and hence, this is a tougher baseline to compare to, and illustrates the capability of our algorithm much more clearly.}. The importance of these tests is especially marked for \textit{T1}, which has no measure of current drawn. Since in general the discharge may not follow a linear decay \cite{POWER_002}, the resulting profiling (from 100\% to 80\% over a period of 5 days) supported our decision to extrapolate from a linear idle consumption.
Figure \ref{FIG_IDLE_DISCHARGE_V} shows the evolution of the voltage with time as the devices discharge slowly (typically over a number of days). For brevity's sake, only the discharge for device \textit{T1} is shown. The charge level shows periodic oscillations but a general linear trend can be inferred (dotted line). The linear estimate was fitted through least-squares of degree 2 and the squared term was actually 6-7 orders of magnitude less than the linear term, meaning that the discharge can be assumed linear. The discharge rates were estimated to be 1.79, 1.06, 5.02 and 2.17 units per hour for \textit{T1}, \textit{T2}, \textit{S1} and \textit{S2} respectively. The rate is higher for smartphones as opposed to tablets (due to the different battery capacities) but also shows a trend towards higher efficiency with the DOZE-feature in Android 6.0.

\subsubsection{Active Battery Consumption}
\begin{figure}[!ht]
	\centering
	\includegraphics[width=\textwidth]{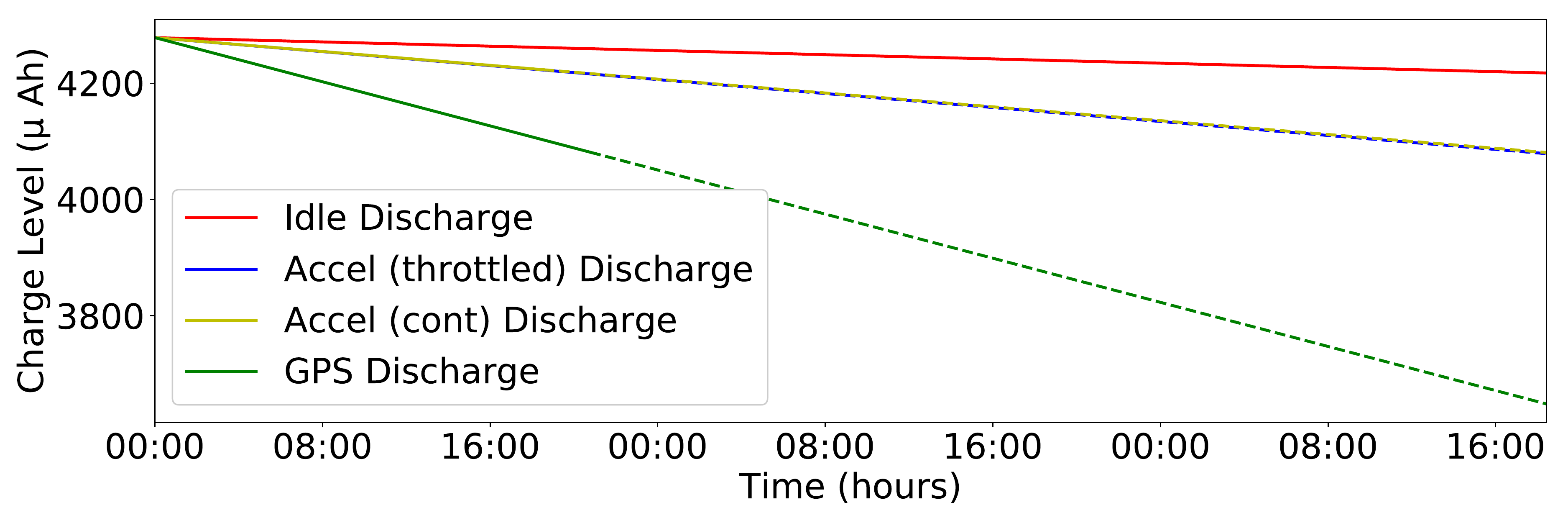}
    \caption{Discharge profile for one of the devices under test (T2). The solid lines are the fitted discharge which is extrapolated linearly (dotted lines).}
    \label{FIG_ACTIVE_DISCHARGE_V}
\end{figure}

We studied power consumption of the GPS (base algorithm), accelerometer and the use of the combined wake-lock by recording and comparing the battery discharge rate, rather than by generating a detailed map of power usage (c.f. \cite{POWER_001}).
The results, tabulated in Table \ref{TAB_DISCHARGE_RATES} and displayed graphically (for \textit{T2} and \textit{S1}) in Fig.\ \ref{FIG_ACTIVE_DISCHARGE_V}, show significant differences between the various algorithms. While turning on the accelerometer consumes an average of three times the idle rate, the GPS increases the rate by an order of magnitude (nine to thirteen times the idle rate).

\begin{table}[!ht]
\centering
\begin{small}
\begin{tabular}{|l|c|c|c|c|}
\hline
\rule{0pt}{3ex}\textbf{Device} & T1 & T2 & S1 & S2 \\
\hline
\hline
\rule{0pt}{3ex} \textbf{\textsl{IDLE}}          &  1.79 &  1.06 &  5.02 &  2.17 \\
\hline
\rule{0pt}{3ex} \textbf{\textsl{Wake-Lock}}     &  6.16 &  --   &  --   &  6.54 \\
\hline
\rule{0pt}{3ex} \textbf{\textsl{Accel. Cont.}}  &  6.71 &  3.22 &  6.70 &  7.42 \\
\hline
\rule{0pt}{3ex} \textbf{\textsl{Accel./Sleep}}  &  6.58 &  3.20 &  --   &  7.22 \\
\hline
\rule{0pt}{3ex} \textbf{\textsl{GPS}}     		& 20.13 & 10.22 & 47.97 & 28.26 \\
\hline
\end{tabular}
\caption{Discharge Rates under various conditions (first-order coefficient of quadratic polynomial)}
\label{TAB_DISCHARGE_RATES}
\end{small}
\end{table}

In the literature there is hardly any consensus as to whether achieving a GPS fix or not affects power consumption, \cite{POWER_006,POWER_001,Prelipcean_2014}.
To study this, a four-way test was set up, with identical devices turned on under the conditions of a clear view of the sky and indoors, as well as in different android location modes utilising either the GPS only or with assistance from Wi-Fi/Cell information. The average measured discharge rates when outdoors  are 20.23 (GPS only) and 22.75 (assisted mode), whilst for indoors rates are 26.37 (GPS only) and 28.87 (assisted mode), or approximately 30\% higher. 
Additionally, assisted location services (which mitigate GPS signal loss) add to further energy consumption, unless the algorithm uses these as another mode to turn the GPS off.

\subsubsection{Motion-Detection FSM Tuning}
The choice of appropriate window-sizes and distance thresholds for the motion-detection FSM were determined using simulated annealing. The five free parameters are:
\begin{inparaenum}[(i)]
\item length of initial sample buffer (1 to 20),
\item motion threshold for the initial buffer  0 to 5),
\item size of second sample buffer (1 to 20),
\item motion threshold for the second buffer (0 to 5), and
\item size of sample-windows over which to average the second buffer (1 to 5).
\end{inparaenum}
The accelerometer traces were used for training the identification algorithm (classifies: motion/no motion) and the parameters optimised using a cost function (Eq \ref{EQ_COST_FUNCTION}) based on  False-Negative Rate (FNR) and False-Positive Rate (FPR), 
number of samples till detection of a True Negative ($N_N$), and number of samples till detection of a True Positive ($N_P$), to regularise the process and avoid overfitting.
\begin{equation}
  C = 12\times FNR + 4\times FPR + 0.02\times N_N +0.04\times N_P \label{EQ_COST_FUNCTION} 
\end{equation}
In line with the needs of the application, the FNR is weighted three times the cost of the FPR (since it is more significant), while cost for sample-delays are two orders of magnitude less than the error rates. The algorithm was 
executed for 10000 epochs, with the experiment repeated multiple times. The optimal parameters indicated a first sample buffer of size 5 samples (about 1s) with a conservatively low-threshold of 0.18 followed by a single sample buffer of size 7, with a threshold of 4.78. In this scenario, the performance on the training data achieved appears in Table \ref{TAB_OPTIMAL_TRAIN} (validation is discussed next).
\begin{table}
\centering
\begin{tabular}{c|c|c}
Predicted \textbackslash Actual & Negative & Positive \\
\hline
\rule{0pt}{3ex} Negative        & \textbf{8104 (96.0\%)} & 295 (7.3\%) \\
\hline
\rule{0pt}{3ex} Positive        & 338 (4.0\%) &  \textbf{3738 (92.7\%)} \\
\end{tabular}
\caption{Optimal Classification Results (training data)}
\label{TAB_OPTIMAL_TRAIN}
\end{table}
Interestingly, the search converges to a low-threshold followed by a high one. This follows from the intuition that the first threshold serves to `feel' the acceleration and hence there is no need to investigate further if there is no motion. Also, counter-intuitively, the second run always converges to a single averaging window rather than multiple small ones.

\subsubsection{Performance Validation}
\label{SSS_PERF_VALID}

\begin{figure}
\centering
\includegraphics[width=\columnwidth]{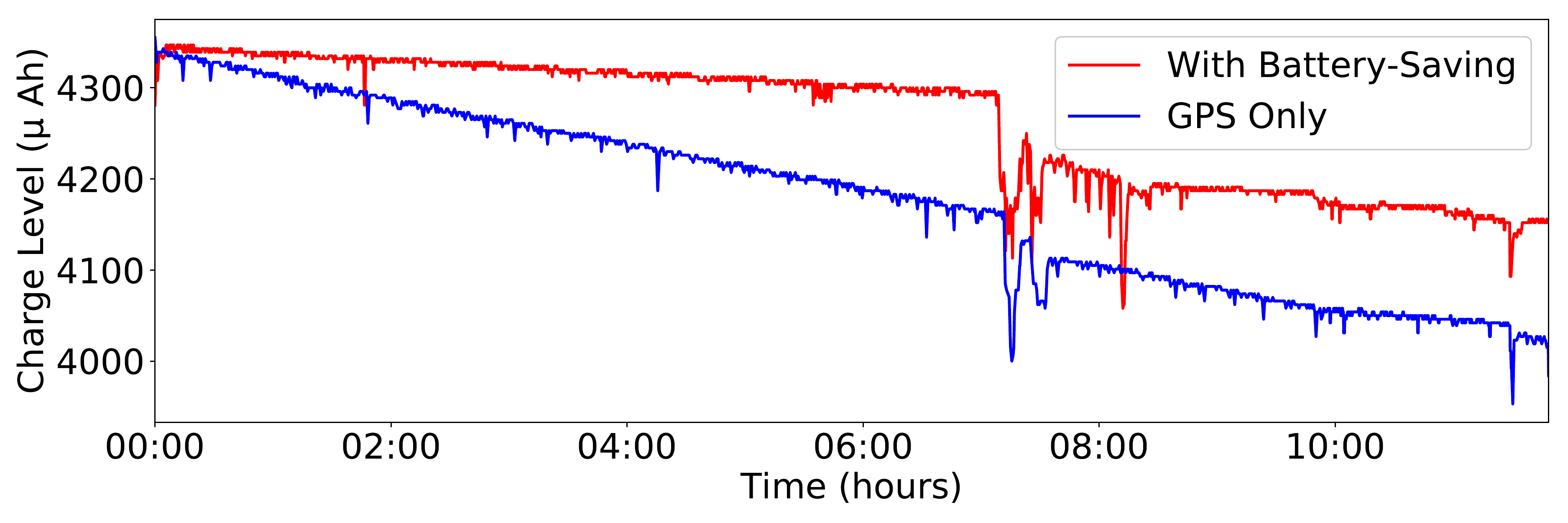}
\caption{Typical Battery Discharge with (red) and without (blue) battery-saving}
\label{FIG_DISCHARGE_WOBS}
\end{figure}

Finally, we ran tests for typical usages throughout a single day with the baseline algorithm and the battery-aware version.
We use these results to demonstrate the efficacy of our system.
In this case, participants were given two identical devices (\textit{S2}), one running the base algorithm and the other the battery-aware adaptation, as they went about their normal day routine.
The device running the battery-aware scheme detected 88\% of all journeys recorded using the base algorithm (i.e. without the Battery-Saving scheme): 9\% of the journeys were clipped at the start or end of the journey and the rest involved a trip on a ferry which was not detected, possibly due to minimal acceleration.
The battery-aware algorithm exhibited savings of between 50\% and 70\% in total battery consumption, compared to the base algorithm (see Fig.\ \ref{FIG_DISCHARGE_WOBS}).
Furthermore  participants who donated their data reported that the version with the battery-saving algorithm is significantly better compared to the one without, in the sense that they were not worried of running out of battery power towards the end of the day, contributing to sustained uptake.

\section{Conclusions}
\label{S_CONCLUSIONS}

In this paper we reported on the development of a smartphone based trip detection and mobility data collection application.
We defined what constitutes a journey and identified a number of issues which may arise in naively using GPS traces.
We described in detail the trip segmentation algorithm, its implementation and its battery-saving schemes, which allowed the phone to track a full day's worth of journeys without the need of  re-charging.
We tuned the free parameters of the algorithms by optimising on captured data, and verified that our technique is able to capture most of the trips in a real-world mixed-mode scenario (97\%, if considering all trips logged) with significant battery savings of up to 70\%.

Future work is required to further fine-tune many of the thresholds with more more in-the-field experiments to verify and understand their effects. 
It is also interesting to look at using map data to enhance the concatenating scheme, especially as regards transport mode changes: other outlets include measuring the bias in the rate of under/over-reporting of the measurement method.


\begin{thebibliography}{10}
\expandafter\ifx\csname url\endcsname\relax
  \def\url#1{\texttt{#1}}\fi
\expandafter\ifx\csname urlprefix\endcsname\relax\def\urlprefix{URL }\fi
\expandafter\ifx\csname href\endcsname\relax
  \def\href#1#2{#2} \def\path#1{#1}\fi

\bibitem{stopher_2015}
P.~R. Stopher, L.~Shen, W.~Liu, A.~Ahmed,
  \href{http://www.sciencedirect.com/science/article/pii/S2352146515003099}{The
  challenge of obtaining ground truth for GPS processing}, Transportation
  Research Procedia 11 (2015) 206 -- 217, transport Survey Methods: Embracing
  Behavioural and Technological Changes Selected contributions from the 10th
  International Conference on Transport Survey Methods 16-21 November 2014,
  Leura, Australia.
\newblock \href {http://dx.doi.org/https://doi.org/10.1016/j.trpro.2015.12.018}
  {\path{doi:https://doi.org/10.1016/j.trpro.2015.12.018}}.

\bibitem{geurs_2015}
K.~T. Geurs, T.~Thomas, M.~Bijlsma, S.~Douhou,
  \href{http://www.sciencedirect.com/science/article/pii/S2352146515003130}{Automatic
  trip and mode detection with move smarter: First results from the Dutch
  mobile mobility panel}, Transportation Research Procedia 11 (2015) 247 --
  262, transport Survey Methods: Embracing Behavioural and Technological
  Changes Selected contributions from the 10th International Conference on
  Transport Survey Methods 16-21 November 2014, Leura, Australia.
\newblock \href {http://dx.doi.org/https://doi.org/10.1016/j.trpro.2015.12.022}
  {\path{doi:https://doi.org/10.1016/j.trpro.2015.12.022}}.

\bibitem{berger_2015}
M.~Berger, M.~Platzer,
  \href{http://www.sciencedirect.com/science/article/pii/S2352146515003142}{Field
  evaluation of the smartphone-based travel behaviour data collection app
  ``smartmo''}, Transportation Research Procedia 11 (2015) 263 -- 279,
  transport Survey Methods: Embracing Behavioural and Technological Changes.
  Selected contributions from the 10th International Conference on Transport
  Survey Methods 16-21 November 2014, Leura, Australia.
\newblock \href {http://dx.doi.org/https://doi.org/10.1016/j.trpro.2015.12.023}
  {\path{doi:https://doi.org/10.1016/j.trpro.2015.12.023}}.

\bibitem{zhao_2015}
F.~Zhao, A.~Ghorpade, F.~C. Pereira, C.~Zegras, M.~Ben-Akiva,
  \href{http://www.sciencedirect.com/science/article/pii/S2352146515003105}{Stop
  detection in smartphone-based travel surveys}, Transportation Research
  Procedia 11 (2015) 218 -- 226, transport Survey Methods: Embracing
  Behavioural and Technological Changes Selected contributions from the 10th
  International Conference on Transport Survey Methods 16-21 November 2014,
  Leura, Australia.
\newblock \href {http://dx.doi.org/https://doi.org/10.1016/j.trpro.2015.12.019}
  {\path{doi:https://doi.org/10.1016/j.trpro.2015.12.019}}.

\bibitem{Prelipcean_2014}
A.~C. Prelipcean, G.~Gidofalvi, Y.~O. Susilo,
  \href{http://www.tandfonline.com/doi/abs/10.1080/17489725.2014.973917}{{Mobility
  Collector}}, Journal of Location Based Services 8~(4) (2014) 229--255.
\newblock \href {http://dx.doi.org/10.1080/17489725.2014.973917}
  {\path{doi:10.1080/17489725.2014.973917}}.

\bibitem{SEGMENT_006}
C.~Cottrill, F.~Pereira, F.~Zhao, I.~Ferreira~Dias, H.~Beng~Lim, M.~Ben-Akiva,
  C.~Zegras, \href{http://dx.doi.org/10.3141/2354-07}{Future mobility survey},
  Transportation Research Record: Journal of the Transportation Research Board
  2354 (2013) 59--67.
\newblock \href {http://dx.doi.org/10.3141/2354-07}
  {\path{doi:10.3141/2354-07}}.

\bibitem{PRELIPCEAN_2018}
A.~C. Prelipcean, Y.~O. Susilo, G.~Gidofalvi,
  \href{http://www.sciencedirect.com/science/article/pii/S2352146518301832}{Collecting
  travel diaries: Current state of the art, best practices, and future research
  directions}, Transportation Research Procedia 32 (2018) 155 -- 166, transport
  Survey Methods in the era of big data:facing the challenges.
\newblock \href {http://dx.doi.org/https://doi.org/10.1016/j.trpro.2018.10.029}
  {\path{doi:https://doi.org/10.1016/j.trpro.2018.10.029}}.

\bibitem{Prelipcean_2018c}
A.~C. Prelipcean, G.~Gidofalvi, Y.~Susilo,
  \href{https://www.sciencedirect.com/science/article/pii/S0198971517305240}{Meili:
  A travel diary collection, annotation and automation system}, Computers,
  Environment and Urban Systems 70~(July 2018) (2018) 24--34, qC 20180605.
\newblock \href {http://dx.doi.org/10.1016/j.compenvurbsys.2018.01.011}
  {\path{doi:10.1016/j.compenvurbsys.2018.01.011}}.

\bibitem{Zheng_2008}
Y.~Zheng, L.~Liu, L.~Wang, X.~Xie,
  \href{http://doi.acm.org/10.1145/1367497.1367532}{Learning transportation
  mode from raw gps data for geographic applications on the web}, in:
  Proceedings of the 17th International Conference on World Wide Web, WWW '08,
  ACM, New York, NY, USA, 2008, pp. 247--256.
\newblock \href {http://dx.doi.org/10.1145/1367497.1367532}
  {\path{doi:10.1145/1367497.1367532}}.

\bibitem{Stenneth_2011}
L.~Stenneth, O.~Wolfson, P.~S. Yu, B.~Xu,
  \href{http://doi.acm.org/10.1145/2093973.2093982}{Transportation mode
  detection using mobile phones and gis information}, in: Proceedings of the
  19th ACM SIGSPATIAL International Conference on Advances in Geographic
  Information Systems, GIS '11, ACM, New York, NY, USA, 2011, pp. 54--63.
\newblock \href {http://dx.doi.org/10.1145/2093973.2093982}
  {\path{doi:10.1145/2093973.2093982}}.

\bibitem{Zhang_2011}
L.~Zhang, S.~Dalyot, D.~Eggert, M.~Sester, Multi-stage approach to travel-mode
  segmentation and classification of gps traces, ISPRS - International Archives
  of the Photogrammetry, Remote Sensing and Spatial Information Sciences
  XXXVIII-4/W25.
\newblock \href {http://dx.doi.org/10.5194/isprsarchives-XXXVIII-4-W25-87-2011}
  {\path{doi:10.5194/isprsarchives-XXXVIII-4-W25-87-2011}}.

\bibitem{Lee_2012}
C.~Lee, G.~Yoon, D.~Han, \href{http://doi.acm.org/10.1145/2442810.2442826}{A
  context-based energy optimization algorithm for periodic localization in
  smartphones}, in: Proceedings of the First ACM SIGSPATIAL International
  Workshop on Mobile Geographic Information Systems, MobiGIS '12, ACM, New
  York, NY, USA, 2012, pp. 86--92.
\newblock \href {http://dx.doi.org/10.1145/2442810.2442826}
  {\path{doi:10.1145/2442810.2442826}}.

\bibitem{Biljecki_2013}
F.~Biljecki, H.~Ledoux, P.~van Oosterom, Transportation mode-based segmentation
  and classification of movement trajectories, International Journal of
  Geographical Information Science 27 (2013) 385--407.

\bibitem{Rasmussen_2013}
T.~K. Rasmussen, J.~B. Ingvardson, K.~Halldórsdóttir, O.~A. Nielsen, Using
  wearable gps devices in travel surveys: A case study in the greater
  copenhagen area, in: Proceedings from the Annual Transport Conference at
  Aalborg University, 2013, pp. 1603--9696.

\bibitem{Nitsche_2014}
P.~Nitsche, P.~Widhalm, S.~Breuss, N.~Br\"andle, P.~Maurer,
  \href{http://www.sciencedirect.com/science/article/pii/S0968090X13002325}{Supporting large-scale travel surveys with
smartphones -- A practical approach},
  Transportation Research Part C: Emerging Technologies 43 (2014) 212 --221,
  special Issue with Selected Papers from Transport Research Arena.
\newblock \href {http://dx.doi.org/https://doi.org/10.1016/j.trc.2013.11.005}
  {\path{doi:https://doi.org/10.1016/j.trc.2013.11.005}}.

\bibitem{Safi_2014}
H.~Safi, B.~Assemi, M.~Mesbah, F.~Luis, H.~Mark, Design and implementation of a
  smartphone-based system for personal travel survey: Case study from new
  zealand, in: Presented at Transportation Research Board 94th Annual Meeting,
  2015.

\bibitem{Stenneth_2012}
L.~Stenneth, K.~Thompson, W.~Stone, J.~Alowibdi, Automated transportation
  transfer detection using gps enabled smartphones, in: 2012 15th International
  IEEE Conference on Intelligent Transportation Systems, 2012, pp. 802--807.
\newblock \href {http://dx.doi.org/10.1109/ITSC.2012.6338603}
  {\path{doi:10.1109/ITSC.2012.6338603}}.

\bibitem{Xiao_2015}
G.~Xiao, Z.~Juan, J.~Gao, Inferring trip ends from gps data based on
  smartphones in shanghai, in: Proceedings of the Transportation Research Board
  94th Annual Meeting, 2015, pp. 11--15.

\bibitem{Prelipcean_2016}
A.~Prelipcean, G.~Gidofalvi, Y.~Susilo, Measures of transport mode
  segmentation of trajectories, International Journal of Geographical
  Information Science 30 (2016) 1--22.
\newblock \href {http://dx.doi.org/10.1080/13658816.2015.1137297}
  {\path{doi:10.1080/13658816.2015.1137297}}.

\bibitem{POWER_003}
S.~Barbeau, M.~A. Labrador, A.~Perez, P.~Winters, N.~Georggi, D.~Aguilar,
  R.~Perez, Dynamic management of real-time location data on gps-enabled mobile
  phones, in: 2008 The Second International Conference on Mobile Ubiquitous
  Computing, Systems, Services and Technologies, 2008, pp. 343--348.
\newblock \href {http://dx.doi.org/10.1109/UBICOMM.2008.83}
  {\path{doi:10.1109/UBICOMM.2008.83}}.

\bibitem{POWER_006}
F.~Ben~Abdesslem, A.~Phillips, T.~Henderson,
  \href{http://doi.acm.org/10.1145/1592606.1592621}{Less is more:
  Energy-efficient mobile sensing with senseless}, in: Proceedings of the 1st
  ACM Workshop on Networking, Systems, and Applications for Mobile Handhelds,
  MobiHeld '09, ACM, New York, NY, USA, 2009, pp. 61--62.
\newblock \href {http://dx.doi.org/10.1145/1592606.1592621}
  {\path{doi:10.1145/1592606.1592621}}.

\bibitem{POWER_009}
T.~O. Oshin, S.~Poslad, A.~Ma, Improving the energy-efficiency of gps based
  location sensing smartphone applications, in: 2012 IEEE 11th International
  Conference on Trust, Security and Privacy in Computing and Communications,
  2012, pp. 1698--1705.
\newblock \href {http://dx.doi.org/10.1109/TrustCom.2012.184}
  {\path{doi:10.1109/TrustCom.2012.184}}.

\bibitem{POWER_007}
I.~Constandache, S.~Gaonkar, M.~Sayler, R.~R. Choudhury, L.~Cox, Enloc:
  Energy-efficient localization for mobile phones, in: IEEE INFOCOM 2009, 2009,
  pp. 2716--2720.
\newblock \href {http://dx.doi.org/10.1109/INFCOM.2009.5062218}
  {\path{doi:10.1109/INFCOM.2009.5062218}}.

\bibitem{POWER_010}
Z.~Zhuang, K.-H. Kim, J.~P. Singh,
  \href{http://doi.acm.org/10.1145/1814433.1814464}{Improving energy efficiency
  of location sensing on smartphones}, in: Proceedings of the 8th International
  Conference on Mobile Systems, Applications, and Services, MobiSys '10, ACM,
  New York, NY, USA, 2010, pp. 315--330.
\newblock \href {http://dx.doi.org/10.1145/1814433.1814464}
  {\path{doi:10.1145/1814433.1814464}}.

\bibitem{POWER_011}
J.~Paek, J.~Kim, R.~Govindan,
  \href{http://doi.acm.org/10.1145/1814433.1814463}{Energy-efficient
  rate-adaptive gps-based positioning for smartphones}, in: Proceedings of the
  8th International Conference on Mobile Systems, Applications, and Services,
  MobiSys '10, ACM, New York, NY, USA, 2010, pp. 299--314.
\newblock \href {http://dx.doi.org/10.1145/1814433.1814463}
  {\path{doi:10.1145/1814433.1814463}}.

\bibitem{POWER_012}
D.~H. Kim, Y.~Kim, D.~Estrin, M.~B. Srivastava,
  \href{http://doi.acm.org/10.1145/1869983.1869989}{Sensloc: Sensing everyday
  places and paths using less energy}, in: Proceedings of the 8th ACM
  Conference on Embedded Networked Sensor Systems, SenSys '10, ACM, New York,
  NY, USA, 2010, pp. 43--56.
\newblock \href {http://dx.doi.org/10.1145/1869983.1869989}
  {\path{doi:10.1145/1869983.1869989}}.

\bibitem{HAVERSINE_001}
C.~C. Robusto, \href{http://www.jstor.org/stable/2309088}{The cosine-haversine
  formula}, The American Mathematical Monthly 64~(1) (1957) 38--40.
\newline\urlprefix\url{http://www.jstor.org/stable/2309088}

\bibitem{GPS_2011}
P.~Misra, P.~Enge,
  \href{https://books.google.com.mt/books?id=5WJOywAACAAJ}{Global Positioning System: Signals, Measurements, and Performance}, Ganga-Jamuna Press, 2011.
\newline\urlprefix\url{https://books.google.com.mt/books?id=5WJOywAACAAJ}

\bibitem{POWER_001}
A.~Carroll, G.~Heiser, {An analysis of power consumption in a smartphone}, in:
  USENIX, 2010.

\bibitem{POWER_002}
L.~Hruska, {Smart batteries and lithium ion voltage profiles}, The Twelfth Annual Battery Conference on Applications and Advances (1997) 205--210\href{http://dx.doi.org/10.1109/BCAA.1997.574104}{\path{doi:10.1109/BCAA.1997.574104}}.

\end{thebibliography}
\end{document}